\documentclass[foma;,3p,10pt]{elsarticle}

\usepackage{graphicx}
\usepackage{amssymb}
\usepackage{amsmath}
\usepackage{soul}
\usepackage[dvipsnames]{xcolor}
\usepackage{subfig}

\usepackage[version=4]{mhchem}

\usepackage{fancyvrb}
\usepackage{framed,xcolor}
\colorlet{shadecolor}{blue!20}
\usepackage[frozencache,cachedir=mintedmain]{minted}

\def\vector#1{\mbox{\boldmath $#1$}}

\def\tensor#1{\mathcal #1}

\makeatletter
\def\@author#1{\g@addto@macro\elsauthors{\normalsize%
    \def\baselinestretch{1}%
    \upshape\authorsep#1\unskip\textsuperscript{%
      \ifx\@fnmark\@empty\else\unskip\sep\@fnmark\let\sep=,\fi
      \ifx\@corref\@empty\else\unskip\sep\@corref\let\sep=,\fi
      }%
    \def\authorsep{\unskip,\space}%
    \global\let\@fnmark\@empty
    \global\let\@corref\@empty  
    \global\let\sep\@empty}%
    \@eadauthor={#1}
}
\makeatother

\makeatletter
\def\ps@pprintTitle{%
   \let\@oddhead\@empty
   \let\@evenhead\@empty
   \let\@oddfoot\@empty
   \let\@evenfoot\@oddfoot
}

\begin{document}

\begin{frontmatter}


\title{An integrated heterogeneous computing framework for ensemble simulations of laser-induced ignition}

\author[1]{Kazuki Maeda\corref{cor1}}
\ead{kemaeda@stanford.edu}
\cortext[cor1]{Corresponding author}
\author[2]{Thiago Teixeira}
\ead{thiagoxt@stanford.edu}
\author[1]{Jonathan M. Wang}
\ead{jmwang14@stanford.edu}
\author[3]{Jeffrey M. Hokanson}
\ead{jeho8774@colorado.edu }
\author[4]{Caetano Melone}
\ead{cmelone@stanford.edu}
\author[5]{Mario {Di Renzo}}
\ead{direnzo.mario1@gmail.com }
\author[6]{Steve Jones}
\ead{stevejones@stanford.edu }
\author[1]{Javier Urzay}
\ead{jurzay@stanford.edu }
\author[6]{Gianluca Iaccarino}
\ead{jops@stanford.edu}

\address[1]{Center for Turbulence Research, Stanford University, USA}
\address[2]{Department of Computer Science, Stanford University, USA}
\address[3]{Smead Aerospace Engineering Sciences, University of Colorado at Boulder, USA}
\address[4]{High Performance Computing Center, Stanford University, USA}
\address[5]{Centre Européen de Recherche et de Formation Avancée en Calcul Scientifique, France}
\address[6]{Department of Mechanical Engineering, Stanford University, Stanford University, USA}

\begin{abstract}
An integrated computational framework is introduced to study complex engineering systems through physics-based ensemble simulations on heterogeneous supercomputers. The framework is primarily designed for the quantitative assessment of laser-induced ignition in rocket engines. We develop and combine an implicit programming system, a compressible reacting flow solver, and a data generation/management strategy on a robust and portable platform. We systematically present this framework using test problems on a hybrid CPU/GPU machine. Efficiency, scalability, and accuracy of the solver are comprehensively assessed with canonical unit problems. Ensemble data management and autoencoding are demonstrated using a canonical diffusion flame case. Sensitivity analysis of the ignition of a turbulent, gaseous fuel jet is performed using a simplified, three-dimensional model combustor. Our approach unifies computer science, physics and engineering, and data science to realize a cross-disciplinary workflow. The framework is exascale-oriented and can be considered a benchmark for future computational science studies of real-world systems.
\end{abstract}

\end{frontmatter}

\section{Introduction}
Predictive computational science and engineering (CSE) research studies nowadays require extensive software development activities and the integration of cross-disciplinary efforts in computer science, physics and engineering, and data science on high-performance computers with complex machine architectures.
This is in contrast to traditional CSE studies in which simulation workflows have often been designed and deployed among few physicists and/or engineers. Physics solvers have been traditionally written in common languages, such as Fortran and C++, within a standard parallel framework like Message Passing Interface (MPI) \citep{Gropp99}.
Now, state-of-the-art HPC systems are employing increasingly heterogeneous processors, such as hybrid CPU-GPU systems, and complex memory hierarchies to simultaneously achieve performance gain and energy efficiency \citep{Mittal15,Vetter19}. 
The existing programming systems are typically specialized for homogeneous architectures and may not take full advantage of the power of heterogeneous computing without extensive interactions with computer scientists.
The development of efficient programming systems and easy-to-use languages for heterogeneous systems are by themselves active topics of computer science (CS).
At the same time, with the growth of computational power, the traffic and the volume of data generated are skyrocketing, and knowledge extraction requires independent expertise in data science (DS).
Future CSE studies thus naturally involve computer scientists for the development and adaptation of programming systems/languages as well as data scientists for big-data analysis.

Combining and complementing diverse expertise requires dedicated integration activities to facilitate (1)the coherent development of the programming system, and simulation and data analysis tools, (2)design of a parallel physics solver for scalable high-fidelity simulations (3)efficient management of ensemble simulations and big-data, and (4)portability of the framework to various machine environments. Although approaches to overcome these difficulties may depend on the overall goal of the study, the challenge of integration itself will only be prominent in broad CSE studies from now on, as compute systems grow toward exascale. To address the challenge, development of a unified computational framework is crucial to accommodate the cross-disciplinary efforts. In fact, the need for such a framework and road-maps have been envisioned for external aerodynamics simulations \citep{Slotnick14}, and practical realization and deployment are critical matters of investigation.

In this work, we present an integrated computational framework for the prediction of laser-induced ignition of chemical rocket engines on heterogeneous systems.
Laser-induced ignition is an important technology for aerospace propulsion systems, in which laser pulse of $O(10)$ ns duration with a wavelength of $O(1)$ $\mu$m is focused into a $O(10-100)$ $\mu$m spot in the rocket combustor, delivering a power density of $O(10-100)$ MWh/m$^2$ and breaking down the gaseous fuel/oxidizer mixture \citep{Phuoc06,O16}. Laser is re-usable and non-intrusive. These features are especially advantageous for igniting re-startable upper stage engines and reaction control thrusters of spacecraft. The technology is also applicable to other systems like internal combustion engines, gas turbine engines, and supersonic combustion ramjets (scramjets) \citep{Brieschenk13,wang2021cnf}.
For further reliable use of the technology, the accurate prediction of key quantities is crucial, including the sensitivity of ignition to local flow conditions, the thermal and mechanical loads on the system, and the ignition uncertainty under low-pressure in high altitude. Meanwhile, rocket ignition is an abrupt and extreme event involving multi-scale, multi-physics interactions of turbulent reacting flows in the space environment, and experiments are limited. Computations of these scenarios are critical to provide insights and confidence in the deployment of this technology.

\begin{figure}[t!]
    \centering
    \includegraphics[width=164mm,trim=20mm 50mm 5mm 20mm, clip]{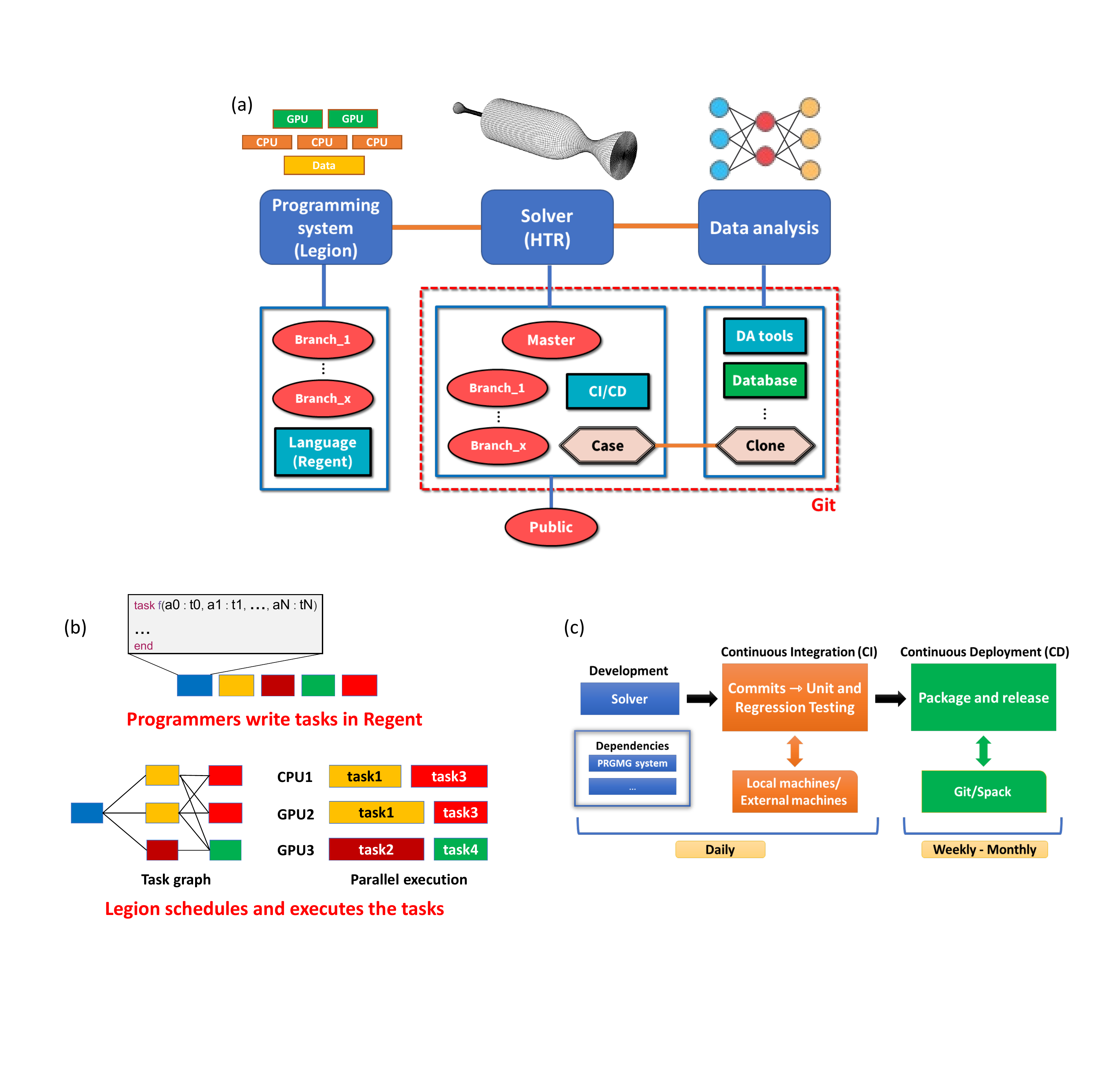}
    \caption{(a) Schematic of the computational framework including three containers: Programming system container, solver container, and data analysis container. (b) Schematic of the Legion/Regent workflow. (c) Schematic of the CI/CD strategy and solver release pipeline.}
    \label{fig1}
\end{figure}

The present contributions are summarized in four areas: (1)integrated workflow including implicit parallel programming system, physics-based simulations, and data-driven surrogates, (2)a scalable solver for high-speed combustion, (3)ensemble data management for uncertainty quantification and sensitivity analysis, and (4)continuous integration and deployment (CI/CD) infrastructure on multiple supercomputer architectures.
The framework accommodates these efforts within interacting containers: programming system, solver, and data analysis (DA) (Fig. \ref{fig1}a).
For the programming system, we employ Legion \citep{Bauer12} for flexible, implicit mapping of tasks on heterogeneous processors.
For the solver, we employ and extend the Hypersonic Task-based Research solver (hereafter HTR) \citep{DiRenzo20,DiRenzo22}, a computational fluid dynamics (CFD) code for high-order accurate simulations of compressible reacting flows in non-canonical geometries. The top-layer of the solver is written in Regent, a high-productivity language for Legion.
The solver development is synchronized with the version-up of Legion. 
The DA container imports a pre-compiled solver executable and use cases from the solver container for smooth database generation and analysis.
CI/CD ensures fault-tolerance and portability of the framework. We describe details of each container, and assess and demonstrate the capabilities of the framework using test problems and example cases which model the rocket ignition on a hybrid CPU/GPU machine.
Our framework not only addresses the complex flow physics for aerospace applications, but also can be considered a benchmark for future CSE studies of real-world systems on the state-of-the-art computing systems. See supplementary information for representative data of the progression of the run-time required for unit tests (Fig. S.1.2).

\section{Results}

\subsection{Programming system}
\subsubsection{Legion programming system}
Figure \ref{fig1}b shows the schematic of our workflow using Legion/Regent.
Legion is a parallel programming system in which apparently sequential programs are automatically parallelized~\citep{Bauer12}.
Programs implemented using the Legion programming model are based on logical regions to describe data organization and to make direct relationships useful for reasoning for locality and independence.
Legion pipelines the execution of the tasks and distributes its runtime execution across the machine. Of central importance to heterogeneous architectures is how tasks are mapped to processors and how physical instances of logical regions are mapped to specific memory units.
The Legion runtime, based on the mapping decisions, automatically copy and transfer the data needed by each task.
The default mapper selects only a single kind of processors, GPUs if available and CPUs otherwise.
In this work, we customize the mapper so that the selection covers multiple kinds of processors.
The simultaneous use of multiple processor kinds can be cumbersome for programmers. This mapping feature of Legion can ease the burden. 

In implicit parallel programming models, the tasks consume data produced by other tasks preceding them in program order. To discover implicit parallelism, Legion performs a dependence analysis to establish a partial legal order on task execution.
This analysis can be made parallel and distributed during runtime by dynamic control replication~\citep{10.1145/3437801.3441587}.

\subsubsection{Regent}
Regent \citep{Slaughter15} is a high-productivity language oriented for high-performance computing using Legion.
Regent features two fundamental abstractions: tasks and logical regions. An optimizing compiler translates Regent programs into Legion, which has native support for tasks and logical regions. Regent simplifies the Legion programming model as many details of programming to the Legion runtime system can be managed statically by the compiler.
Regent is built on top of the Terra with extensive support for meta-programming via multi-stage programming~\citep{10.1145/2491956.2462166}.
These features enable programmers to comprehend and modify programs after a short onboarding period and avoid the need for them to become experts in parallel programming. The Regent programs have fewer lines and are generally simpler than the Legion ones, which improves the readability and maintainability of the code when compared to solvers using, for instance, the MPI framework. Regent provides these advantages without sacrificing scalability.

\subsubsection{Ensemble co-processing}
The custom mapper is particularly useful to perform ensemble simulations on heterogeneous machines. For instance, for multi-fidelity ensemble simulations on CPU/GPU machines that rely for example on grid coarsening to reduce the computational cost of a simulation, one can define high-fidelity (HF), fine resolution samples on GPUs and low-fidelity (LF), coarse resolution samples on CPUs. This configuration allows the efficient use of the heterogeneous processors, unlike typical high-performance applications, which utilize mainly GPUs and leave CPUs mostly idle during simulations. Again, this co-processing can be implemented in the mapper without modifications to the solver. 

\subsection{Solver}
Specific extensions of HTR from its original version \citep{DiRenzo20} have been included for modeling high-speed flows of propellants and their fast chemical reactions induced by laser-induced ignition in a rocket combustor.
The solver employs a high-order conservative finite-difference method for simulation of compressible multi-component Navier-Stokes equations on curvilinear grids. Combustion is modeled by finite-rate chemistry.
In this work we show results using the third-order weighted essentially non-oscillatory scheme (WENO3-Z) \citep{Borges08} and the sixth-order targeted essentially non-oscillatory scheme (TENO6) \citep{Fu16} for spatial discretization of the inviscid fluxes. The viscous flux is computed using the second order central difference scheme. The fluxes on curvilinear coordinates are mapped onto Cartesian coordinates through covariant transformation. This approach enables direct applications of standard flux splitting and reconstruction strategies for Cartesian grids for non-canonical geometries with minimal modifications, without sacrificing discrete conservation and parallel efficiency \citep{Vinokur74,Pirozzoli11}.
The WENO and TENO schemes can induce an appropriate amount of numerical dissipation where required (e.g., near shocks), while maintain high-order accuracy in smooth fields, enabling stable simulations of high-speed flows of our focus. Various other schemes and models can be accommodated within this framework.

\subsection{Data analysis}
\subsubsection{Shared case}
The DA and solver containers are managed under a common git version control system \citep{Chacon14} for the smooth transfer of the use cases (\ref{fig1}a).
The DA container stores branches corresponding to a specific test case a set of minimal solver scripts and an executable, a python script ({\small\verb+Makeinput.py+}), and a json input file ({\small\verb+input.json+}). The executable is pre-compiled in the solver container and the branches are pushed by the developers to the DA container. 
This mechanism allows the users (data scientists) to perform simulations and data analysis using the latest features of the solver before they are merged with the public version.
For usability, the procedure for simulation is kept minimal.

\subsection{CI/CD} 
Figure \ref{fig1}c shows the schematic of the CI/CD.
Multiple developers contribute to the solver's source code to modify and add features on a daily basis. The daily changes in dependencies including Legion can require modifications in the code.
Beyond the activities within the framework, external factors may impact the solver's functionality.
Major factors are updates in the system's environment, including GPU architectures, vendors, networking capabilities, and dependent libraries, which may independently occur in different machines.
Manually ensuring and maintaining the functionality is thus impractical.
The continuous integration (CI) overcomes this challenge.
On a nightly basis, several checks are performed on various systems. Successful compilation of the software is first verified, followed by unit and regression tests, which allow a granular view into the solver's performance on each machine and the automated detection of unexpected errors. Representative data of unit test are show in supplementary Fig. S.1.2.

After new commits are tested by the CI system, they are merged into a master branch.
This branch is deployed through the Spack package manager \citep{spack} to simplify the installation of the software stack. When merges are accumulated, the branch can be released as an update for the open-source version.

\subsection{Ensemble simulation}
Ensemble simulation is critical ingredient to address the realistic engineering questions relevant to optimization, uncertainty quantification, reliability analysis, etc.
For instance, to analyze the sensitivity of the ignition to the laser energy, it is not sufficient to consider only a single simulation; multiple simulations must be performed with varying parameters.
We design the overall computational framework for ensemble simulations and for the analysis of resulting data, rather than running a single, massive simulation to achieve the peak performance.

\section{Solver Performance Analysis}
\subsection{Accuracy}
\begin{figure}[t!]
    \centering
   \includegraphics[width=140mm,trim=0 0mm 0 0mm, clip]{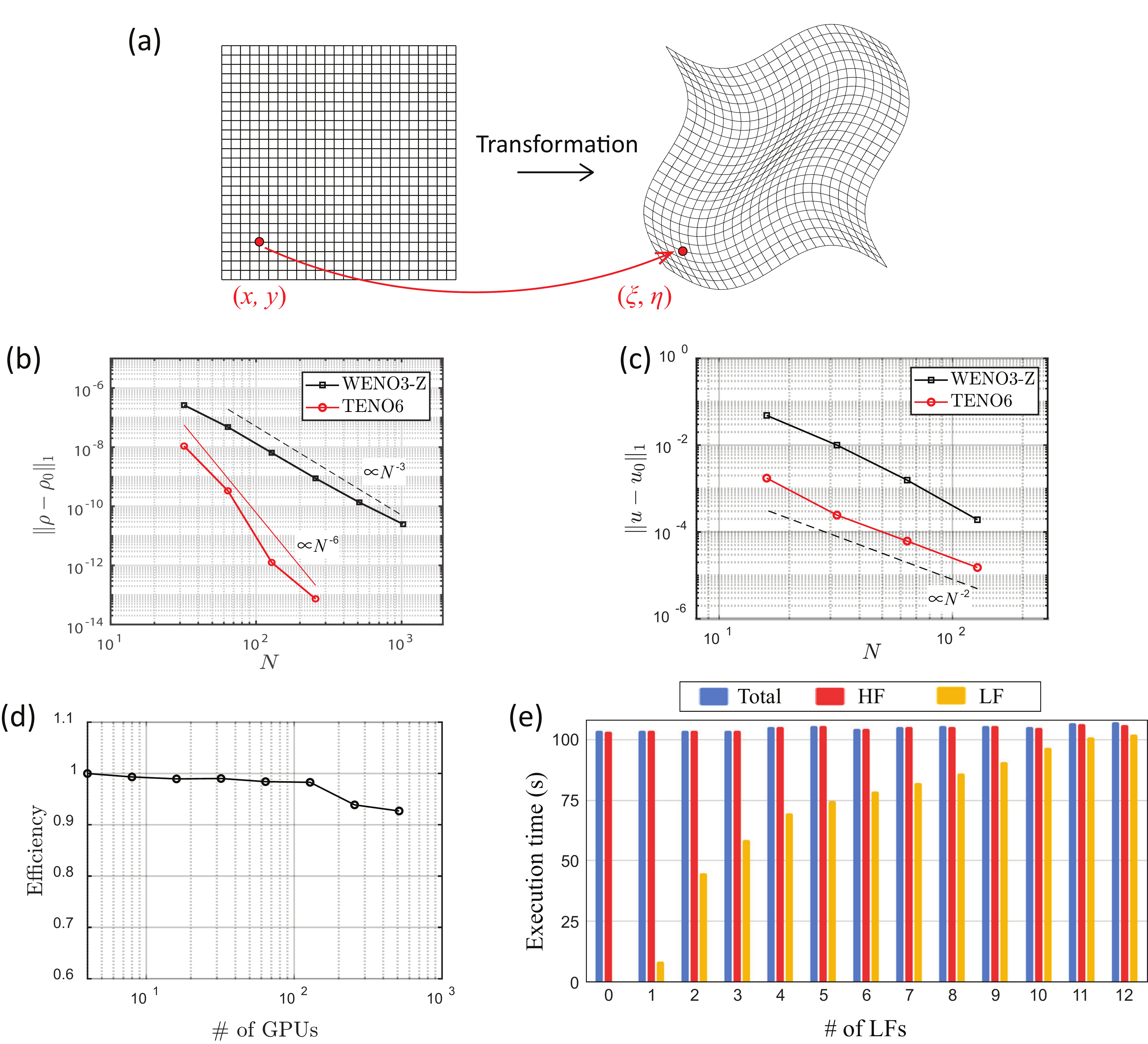}
    \caption{(a) The original mesh and skewed mesh used for the convergence test of the inviscid vortex advection problem. The grids are highly skewed in the center and corners of the domain. (b)$L_1$ norm of the density as a function of $N$. Results of WENO3-Z and TENO6 as well as reference slopes of the third- and sixth order convergence are shown.
    (c) $L_1$ norm of the velocity as a function of $N$. Results of WENO3-Z and TENO6 as well as reference slope of the second order convergence are shown.
    (d) Weak scaling plot obtained on GPUs of Lassen at Lawrence Livermore National Labolatory.
    (e) Execution time of bi-fidelity ensembles as a function of the number of low-fidelity samples. In each ensemble, a high-fidelity (HF) sample is processed on GPUs and multiple low-fidelity (LF) samples are concurrently processed on CPUs within the same nodes. The total time as well as the time required for the HF sample and that of the LF samples are shown.}
    \label{fig2}
\end{figure}
We assess the grid convergence of the numerical solutions in two problems.
The first is the advection of an inviscid vortex for testing the TENO/WENO schemes \citep{Balsara00}.
The second is the Taylor-Green vortex for testing the overall accuracy for simulation of viscous flows.
Skewed curvilinear grids are used by transforming $N\times N$ Cartesian girds, shown in Fig. \ref{fig2}a. Realistic simulations typically avoid using such highly skewed grids for numerical stability. The present grids serve as challenging examples.
Figure \ref{fig2}b shows the $L_1$ error norm of the density in the first problem.
The results indicate high-order convergence with the rates expected by WENO3-Z/TENO6.
Figure \ref{fig2}c shows the $L_1$ error norm of the horizontal velocity in the second problem.
The convergence rate is second-order for both schemes, which can be explained by the discretization of the viscous term. The error is an order of magnitude greater with the WENO3-Z compared to TENO6, at all $N$.
Overall, these results confirm the solver accuracy.
Although TENO6 is globally more accurate than WENO3-Z in these problems, we find that WENO3-Z can be more robust in reacting flow problems that we tested on coarse grids without sub-grid scale modeling.

\subsection{Scalability}
\subsubsection{Weak scaling analysis}
Figure \ref{fig2}d shows the efficiency as a function of the number of GPUs in our weak scaling analysis. Up to 128 GPUs the efficiency is nearly unity and then drops to 0.9 at larger numbers of GPUs. We consider that this step change is due to the saturation in the network communication of the machine \citep{Torres19}.
Overall, the results show the high scalability of the solver.
Supplementary Fig. S.1. shows favorable weak scaling using CPUs as well as on another GPU-CPU machine.

\subsubsection{Ensemble co-processing on GPUs and CPUs}
To assess CPU/GPU co-processing, we consider ensembles, each of which consists of a single HF sample mapped on GPUs and various numbers of LF samples mapped on CPUs in common nodes. Both samples simulate the same physical problem. The HF sample has a greater resolution than the LF sample, making the HF sample more computationally expensive to simulate the same timescale of physics.

We assess the effect of the number of LF samples on the total execution time of the ensembles.
The total execution time is dominated by the execution time of the HF sample, regardless of the number of LF samples (Fig. \ref{fig2}e). The cumulative execution time for the LF samples monotonically increases with the number of the LF samples.
This increase is expected since tasks for each LF sample are queued while previous samples are executed.
Overall, the multiple LF samples are executed on CPUs concurrently with the HF sample on GPUs without deteriorating the total execution time, proving the capability of Legion in efficiently utilizing the heterogeneous resource. A representative task graph for each sample and run-time profiling of ensembles are show in supplementary material (Fig. S.1.3 and S.1.4).

\section{Application examples}
\subsection{Ignition of a diffusion flame}\label{sec:2D-flame}
\subsubsection{Case description}
\begin{figure}[t!]
    \centering
    \includegraphics[width=164mm, trim=0 00mm 0 0mm, clip]{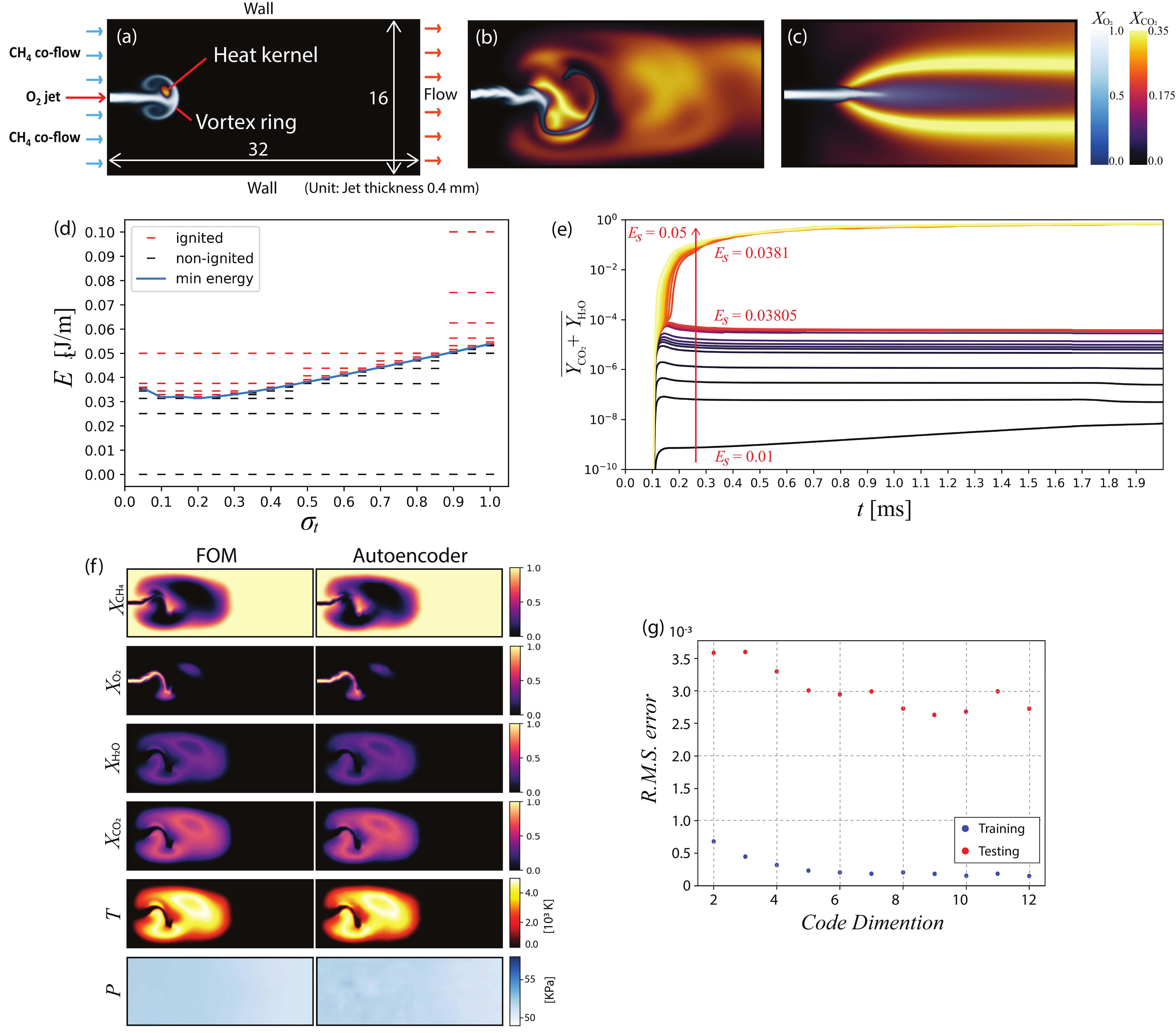}
    \caption{(a-c) Schematic of the problem and snapshots from a representative simulation. (d) Minimum laser energy for successful ignition for various values of the laser duration, obtained from the Bi-section search of the database. (e) Evolution of the molar fraction of the reaction products in the domain for various values of the laser energy, with a fixed laser location. (f) Comparisons of the flow fields of the full-order model and those of the autoencoder. (g) The root-mean-square (R.M.S.) field error of the autoencoder for training and testing as a function of the code dimension (number of the latent variables).}
    \label{fig3}
\end{figure}
The first example case is the ignition of a two-dimensional, non-premixed laminar diffusion flame. This case is a reduced-surrogate capturing the essential physics of laser-induced ignition to assist developing the DA-tools.
Figure \ref{fig3}a-e show temporal snapshots of the field of a representative simulation.
A gaseous \ce{O2} jet is injected into the domain filled with quiescent gaseous \ce{CH4} with the Mach number (Ma) of 0.1 and the Reynolds number (Re) of $400$ with a slow \ce{CH4} co-flow. The right boundary is open.
Shortly afterward, the jet forms a vortex ring (Fig. \ref{fig3}a).
The energy is deposited inside the ring with the modeled-laser. This location is favorable for ignition because the fuel and oxidizer gases are mutually entrained in the vortex and the mixing is locally enhanced.
After the energy deposition, the reaction zone spreads and propagates downstream (Fig. \ref{fig3}b). Eventually, a steady diffusion flame is formed (Fig. \ref{fig3}c).

\subsubsection{Ensemble and data analysis\label{sec:use:UQ}}
For ensemble-based analysis, we construct a database and identify the minimum laser energy for successful ignition.
We fix the location and timing of the energy deposition, and vary the laser energy, $E_s$, and the laser duration in the unit of jet thickness devided by the jet velocity, $\sigma_t$, in $283$ samples. 
For each $\sigma_t$, the bisection line search are performed to determine the minimum energy.
Fireworks \cite{fireworks} is employed to manage/schedule the ensemble.

The minimum energy has a convex profile against $\sigma_t$ with the global minimum of $E_s=0.0313$ J/m at $\sigma_t=0.2$ (Fig. \ref{fig3}d).
The convexity can be explained by the competition between the advection and the wave radiation, both of which are controlled by $\sigma_t$. With increasing $\sigma_t$, the jet is advected in a longer distance during the energy deposition. The laser energy becomes more smeared over the jet and the maximum local energy density decreases. The energy can be deposited outside of the mixture zone. Therefore, with increasing $\sigma_t$, the greater energy is likely to be required for ignition, explaining the increase in the minimum energy at $\sigma_t:\sigma_t\in[0.2,1.0]$.
Meanwhile, with decreasing $\sigma_t$, the energy deposition can cause the radiation of pressure waves from the deposition site. This is because the timescale of the deposition becomes close to that of the acoustic timescale of the flow: $\sigma_s/\sigma_t\sim c_s$, where $c_s$ is the sound speed of the gaseous mixture. The short-time heating excites rapid expansion of the fluid which generates the wave. A part of the deposited energy is carried by this wave rather than the chemical reaction. Therefore the more energy is required for the ignition in the limit of small $\sigma_t$, explaining the negative slope of the minimum energy at $\sigma_t:\sigma_t<0.2$. This compressible flow physics is absent in the limit of large $\sigma_t$.

To assess the sensitivity near the critical point, Figure \ref{fig3}e shows the evolution of the reaction products molar fraction in the domain for samples with various $E_s$ near the threshold with a fixed $\sigma_s: \sigma_s=0.5$.
For $E_s\ge0.38089$, the mass fraction monotonically grows and reaches close to unity, indicating the formation of sustained flame, while for $E_s<0.38089$ the mass fraction remains small and without flame.
The plot indicates that the flame remains once ignited, thus the success/failure of ignition can be judged at early stages.
This clear transition threshold may not hold for realistic rockets in which turbulence-induced fluctuations of the fuel jet can induce stochasticity in the ignition threshold.
Nevertheless, the simplified model can be useful for designing more complex numerical experiments (e.g., Section \ref{s:comb}).

\subsubsection{Autoencoder}
A critical application of the simulation database is machine learning to develop inexpensive surrogates.
The minimum energy database is suitable for this task as the bisection search concentrates samples in the transition region between ignited and non-ignited samples.
To this end, we build a convolutional autoencoder to represent the state of the simulation following~\citep{Lee20}.
Figure~\ref{fig3}f compares the contours of flow quantities in the original data with 1125000 degrees of freedoms and those compressed and reconstructed via 10 latent variables (codes) by the autoencoder. The contours in Figure~\ref{fig3}g shows the root-mean-square field errors for training and testing as a function of the code dimension up to 12. As expected, the error decreases with increasing the code dimension. Both errors are $O(10^{-3})$.
The autoencoder can thus provide a reasonable accuracy of reconstruction with drastic compression. The latent variables vary smoothly in time and with varying laser energy (Fig. S2.2 in supplementary information).

\subsection{Ignition of a model combustor}
\label{s:comb}
\subsubsection{Case description}
\begin{figure}[t!]
    \centering
\includegraphics[width=162mm,trim=0mm 0mm 0mm 0mm, clip]{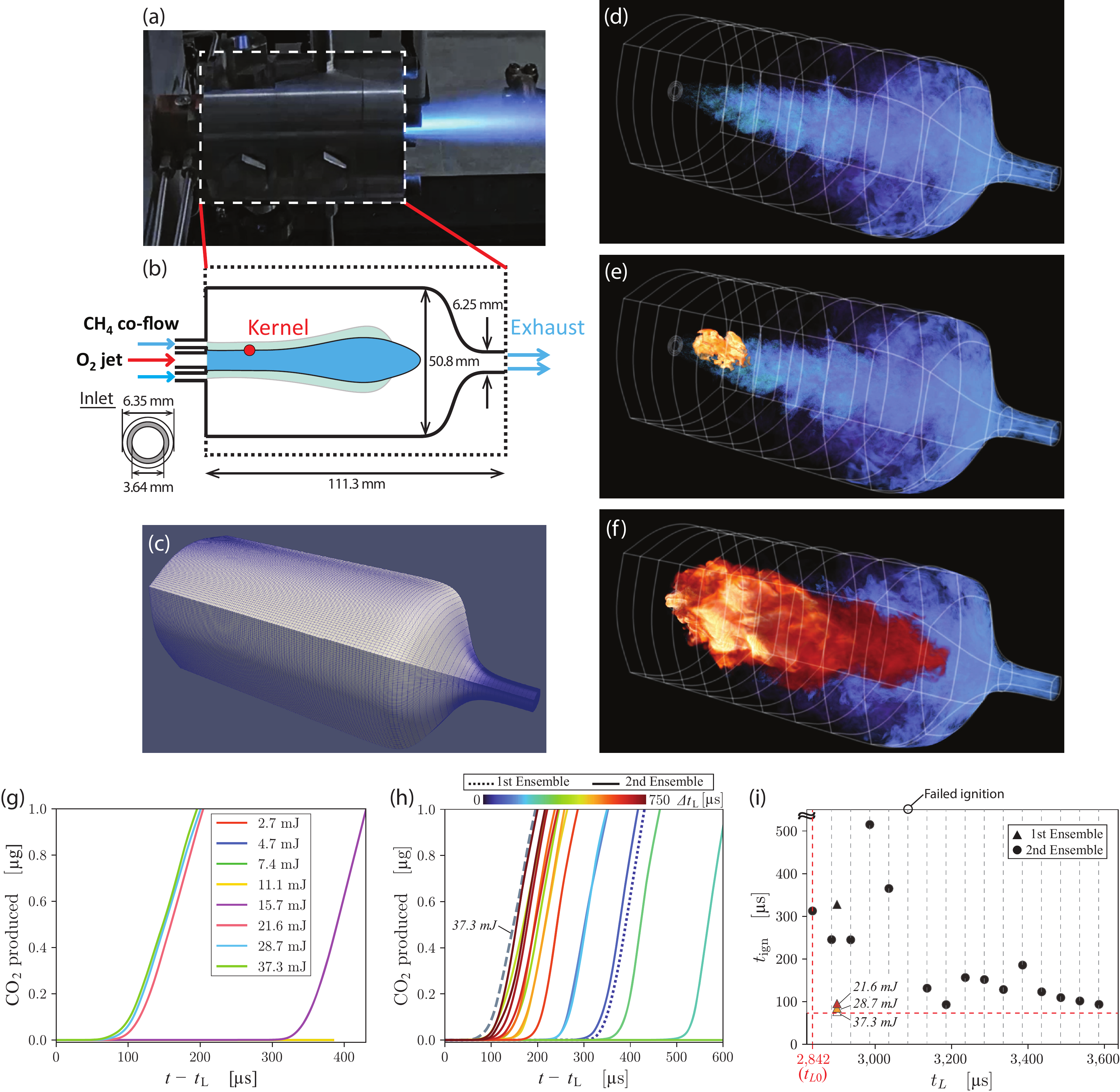}
    \caption{(a) Model combustor under operation in the experiment, courtesy of Carson Slabaugh at Purdue University. (b) Schematic of the combustion chamber. (c) Grid used for the combustor simulations. (d-f) Volume rendering of the ignition kernel (red) and the \ce{CH4} (blue) in a representative simulation at $t$=2906, 3001, and 3246 $\mu$s$^{-1}$, where the laser is deployed in the first image. (d) right before the energy deposition, (e) shortly after the deposition, and (f) during the flame evolution. (g) Time trace of total \ce{CO2} produced for an ensemble of 8 simulations (named as E-1) with varying laser energies: $E_s\in[2.7, 37.3]$ mJ. (h) Time trace of total \ce{CO2} produced for the sample with $E_s=37.3$ mJ in E-1 and that for an ensemble of 16 simulations (named as E-2) with $E_s=15.7$ mJ, where the timing of laser deployment $t_L$ is incremented by $50\,\mu$s between each simulation. The line plots are colored by corresponding $t_L$ subtracted by that of the first sample in E-2: $\Delta t_{L}=t_L-t_{L0}$. $E_s=15.7$ mJ unless noted. Dotted lines correspond to samples from the first ensemble.
    (i) Scatter plot of the ignition delay, $t_{ign}$, against $t_L$, for samples obtained from both ensembles. $E_s=15.7$ mJ unless noted.}
    \label{fig4}
\end{figure}
The second case is a rocket combustor ignition. This case models a real-world companion experiment, although its details are omitted here.
Figures \ref{fig4}a and b show a photograph and schematic, respectively, of the combustor.
The injection consists of an inner, sonic jet ($\text{Ma}=1$) of gaseous \ce{O2} with a subsonic co-flow ($\text{Ma}=0.44$) of gaseous \ce{CH4}, with the constant mass flow rate.
The reactants are injected to pressurize the chamber for a sufficient time for a quasi-stationary-state to be reached (method). The laser is then deployed near the non-premixed \ce{CH4}/\ce{O2} turbulent shear layer. 
Figure \ref{fig4}c shows the mesh used to model the chamber and the nozzle.
Figure \ref{fig4}d-f shows the volume rendering of the flame and the \ce{CH4} mole fraction before (Fig. \ref{fig4}d) and after (Fig.\ref{fig4}e and f) the laser deployment from an ignited simulation sample.
The nascent flame kernel expands and propagates downstream from the laser focal location (Fig.~\ref{fig4}f).
The flame front becomes wrinkled by turbulence.

\subsubsection{Ensemble simulation}
We demonstrate ensemble simulations using Legion's mapping feature (section \ref{s:shared}).
Unlike the diffusion flame (section~\ref{sec:2D-flame}), the high-speed jet is turbulent and 
instantaneous flow quantities near the laser focal location rapidly fluctuate.
Even with the same $E_s$ and laser deployment time, $t_L$, ignition success may vary trial by trial, depending on the fluctuations.
Moreover, ignition may also depend on the slowly varying mean flow.

For analysis, we consider two ensembles, which will be referred to as E-1 and E-2. All simulation samples use the same pre-ignition flows.
E-1 consists of eight samples with fixed $t_L:t_L=2900$ $\mu$s, with various $E_s$.
E-1 estimates the laser-energy threshold above which the ignition is likely to be successful, regardless of the fluctuations.
E-2 consists of sixteen samples with fixed $E_s$, based on E-1, and with various $t_L$. It evaluates the effect of the fluctuations and the mean-flow evolution on ignition for fixed laser energy.

Figure \ref{fig4}g shows the evolution of the \ce{CO2} mass in the combustor for the samples in E-1 as a function of the post-deployment time, $t-t_L$.
The plot shows a clear distinction between ignition success and failure.
For $E_s\geq21.6$ mJ, the time traces have a similar profile.
The \ce{CO2} production is initiated at $t-t_L\approx100\,\mu$\,s and grows monotonically, indicating ignition success. Increasing $E_s$ above 21.6 \text{mJ} does not significantly alter this delay.
For $E_s\leq11.1$ mJ, the production is negligible, indicating ignition failure. 
For $E_s=15.7$ mJ, the production is delayed until $t-t_L=320$ $\mu$s and then grows at a rate similar to the cases with $E_s>15.7$ mJ.
We also observe qualitatively similar flame growth across the ignited samples (supplementary movie S.4.3).

In E-2, we use $E_s=15.7$\,mJ as the threshold. $t_L$ is varied among samples with an increment of 50 $\mu$s starting at $t_L=2842$ $\mu$s ($t_{L0}$). With this increment, the rapid fluctuations in the pre-ignition flow at the focal area become effectively uncorrelated, shown by the auto-correlation of the probe signals in the jet (supplementary Fig. S.3.1. and S.3.2).
Figure \ref{fig4}h shows the evolution of the \ce{CO2} produced for the $E_s=37.3$ mJ sample from E-1 and 17 samples with $E_s=15.7$ mJ for across both ensembles.
The minimum ignition delay is 73 $\mu$s with the highest laser energy ($E_s=37.3$ mJ).
The ignition delay time is spread over a range of 400 $\mu s$, indicating the influence of the flow fluctuations.

Figure \ref{fig4}h further shows that later deployment tends to reduce ignition delay.
Figure \ref{fig4}i shows a scatter plot of the ignition delay time $t_{ign}$ as a function of $t_L$, for samples in both ensembles.
For samples in E-2, the plot shows negative correlation of $t_{ign}$ against $t_L$. For $t_L<3100$ $\mu$s and $E_s=15.7$ mJ, $t_{ign}\ge200$ $\mu$s with relatively large variation. The sixth sample ($t_L=3092$ $\mu$s) is not ignited. For $t_L\ge3100$ $\mu$s, the ignition time is less than $200$ $\mu$s.  For the last five samples, $t_{ign}$ monotonically decreases near the minimum delay.
This stabilization at large $t_{L}$ can be due to the slow increase in the \ce{CH4} molar fraction and pressure and decrease in the streamwise velocity (Fig. S6) of the mean flow, which can be favorable for ignition.
The ensembles therefore quantify the dependence of ignition success on $E_s$ and the sensitivity of $t_{ign}$ to both the turbulent fluctuations and the mean-flow evolution.

\section{Discussion}
The framework presented in this paper is flexible to enable independent scientific investigations in each container as well as mutual feedback.
In the CS container, the Legion mapper can be extended to map tasks for data-driven reduced-order models, which are developed in the DA container, on CPUs so that the models are concurrently simulated with high-fidelity solver samples on GPUs within a single ensemble, without additional expense.
The implementation of new physical models and numerical schemes in the solver can often be done by adding new Regent task modules without interfering with mapping and parallelism.
Moreover, various tools and APIs which are supported by Legion can be accommodated in the framework and mapped with the current solver on heterogeneous processors.
The framework can be ported to next-generation, exascale machines with complex architectures and be of use for the prediction of various physics and engineering systems.

Detailed physics of ignition is out of the scope of this paper. Nevertheless, Our demonstration supports the use of the framework for future uncertainty quantification and sensitivity analysis of rocket ignition.
Identifying the earliest possible timing of the laser deployment for successful ignition is important since late ignition may cause, for instance, ignition overpressure \citep{Manfletti14,Manfletti14b}. Further analysis may require addressing a wider parameter space with a greater number of simulation samples. The use of detailed chemistry and/or turbulence models may improve the accuracy of prediction.

\section{Methods}
\subsection{Shared case}
\label{s:shared}
Here we briefly describe the scripts/procedure for executing the shared case.
First, users generate initial conditions.
To do so, users specify designated parameters (e.g., grid resolution) in the input file and execute the python script.
Example entries of the input file for the diffusion flame case is shown in an appendix for reference.
For ensemble simulation, input files for multiple samples should be executed to generate corresponding initial conditions.
\begin{minted}[fontsize=\scriptsize,bgcolor=lightgray]{bash}
$ MakeInput.pyc -input input0.json
$ MakeInput.pyc -input input1.json
...
\end{minted}
As a result, a directory containing the initial condition (initial flow field read by the solver) and a parameter file ({\small\verb+run0+}, {\small\verb+run1+}) are generated.
Now, an ensemble simulation can be performed as follows.\\
\begin{minted}[fontsize=\scriptsize,bgcolor=lightgray]{bash}
$ QUEUE="queue" [Solver directory]/prometeo.sh -i run0.json -i run1.json ... -o [output directory]
\end{minted}

Output files (e.g., hdf5 files) from samples corresponding to the parameter files are placed in sub-directories under the output directory with a corresponding index ({\small\verb+sample0+}, {\small\verb+sample1+}...).

For multifidelity ensemble simulation, users can pass the parameter files with the low-fidelity option \verb+-lp+.\\
\begin{minted}[fontsize=\scriptsize,bgcolor=lightgray]{bash}
$ QUEUE="queue" [Solver directory]/prometeo.sh -i run0.json -o [output directory] -lp run1.json -lp run2.json ...
\end{minted}
Here, a sample of {\small\verb+run0.json+} runs on GPUs and samples of {\small\verb+run1.json+} and {\small\verb+run2.json+} simultaneously run on CPUs.
The generated data can readily be processed by arbitrary algorithms/softwares suitable on the machine environment to perform data analysis.

\subsection{CI/CD}
For the CI/CD, an automated mechanism to test changes to the solver is built on top of GitLab Runner \citep{runner}. The verification using the unit/regression tests are performed on local systems (Stanford HPCC) as well as external clusters.

\subsection{Solver}
The original version of the solver was developed for simulation of turbulent, multi-species, dissociating gas flows \citep{DiRenzo20}.
The solver has continuously been developed in our framework \citep{DiRenzo22}.
The open-source version of the solver is made available through the public repository of the Stanford High-performance computing center \citep{hpcc}.

Notably, the Legion/Regent framework was employed in Soleil-X \citep{Torres19}, a multi-physics compressible Navier-Stokes solver for radiative heat-transfer in particle-laden flows. HTR partially adapts Soleil-X's code structure for the top-level task organization and case setups.

\paragraph{Governing equations}
We formulate the dynamics of compressible reactive gaseous mixtures using the multi-component Navier-Stokes equations with a chemical source term.
\begin{align}
    &\frac{\partial\rho Y_s}{\partial t}+\nabla\cdot(\rho \vector{u} Y_s)=-\nabla(\rho Y_s\vector{V}_{s})+\dot{\omega}_s,\\
    &\frac{\partial\rho\vector{u}}{\partial t}+\nabla\cdot(\rho\vector{u}\vector{u}+p\tensor{I})=\nabla\cdot\tau,\\
    &\frac{\partial E}{\partial t}+\nabla\cdot( E\vector{u}+p\vector{u})=-\nabla\cdot\vector{q}+\nabla\cdot(\tau\cdot\vector{u})+q_L,
\end{align}
where $\rho$, $\vector{u}$, $E$ are the density, velocity, and the total energy of the mixture.
$Y_s$ is the mass fraction of the $s$-th species. 
$p$ is the pressure.
The set of equations is closed by the ideal gas equation of state
\begin{equation}
    p=\rho RT\sum_s\frac{Y_s}{W_s},
\end{equation}
where $W_s$ is the molar mass of species $s$, and $R$ is the gas constant.
The dynamic viscosity of the mixture is evaluated using Wilke’s rule \citep{Wilke50}.
The specific heat capacities at constant pressure are computed using the nine-coefficient NASA polynomials \citep{Mcbride02}.
The heat flux is expressed as
\begin{equation}
    \vector{q}=-\lambda \nabla T+\sum_s(\rho Y_s \vector{V}_sh_s),
\end{equation}
where $\lambda$ is the mixture thermal conductivity and $h_s$ is the enthalpy of species $s$.
$\vector{V}_s$ is the mass diffusion velocity of species $s$.
\begin{equation}
     \vector{V}_s=-D_s\frac{\nabla X_s}{X_s}+\vector{u}_c,
\end{equation}
where $X_s$ is the mole fraction and $D_s$ is the diffusivity of species $s$.
The diffusivity is computed as a function of the mixture composition and temperature \citep{Bird06} with binary diffusivity \citep{Hirschfelder64}. 
$\vector{u}_c$ is the correction velocity to enforce a zero net diffusion flux.
$q_L$ is the external energy (heat) source modeling the laser deposition.
The definition of $q_L$ is case specific.
The expressions of $q_L$ used in the two applications examples are respectively described in Section 6.3.1 and 6.4.1.
$\dot{\omega}$ is the chemical source term.
Further descriptions about the mixture properties and the evaluation of the chemical source term can be found in \citep{DiRenzo20}.

\paragraph{Combustion chemistry}
The reaction of gaseous methane-oxygen mixtures is modeled using a finite-rate chemistry based on Arrhenius reactions.
The framework supports mechanisms including the one-step global mechanism \citep{CH41}, 12-species reduced mechanism \citep{Xu21}, and the 30 species skeletal mechanism \citep{Lu08}.
Details of these mechanisms are omitted for brevity. The accuracy of the model is, in general, enhanced with the number of the species treated in the mechanisms in expense of computational cost. The choice depends on the intended cost and accuracy of simulations, and is left to the users. Various other mechanisms can also be implemented and used.

\subsection{Numerical method}
To treat general stationary curvilinear coordinates, we first define the governing equations on $x$-$y$-$z$ Cartesian coordinates.
\begin{align}
    \frac{\partial\vector{U}}{\partial t}
    +\frac{\partial\vector{F}^{a}}{\partial x}
    +\frac{\partial\vector{G}^{a}}{\partial y}
    +\frac{\partial\vector{H}^{a}}{\partial z}
    =
     \frac{\partial\vector{F}^{d}}{\partial x}
    +\frac{\partial\vector{G}^{d}}{\partial y}
    +\frac{\partial\vector{H}^{d}}{\partial z}
    +\vector{S},
\end{align}
where $\vector{U}$ is the conservative variable, $\vector{F}$, $\vector{G}$, and $\vector{H}$ are the fluxes in the $x$, $y$, and $z$ directions, respectively. The superscripts $(\cdot)^a$ and $(\cdot)^d$ denote advection (invsicid) and diffusive components of the fluxes, respectively. $\vector{S}$ is the source term.
We then perform transformation in $x$-$y$-$z$ coordinates onto Cartesian grids in $\xi$-$\eta$-$\zeta$ coordinates.
The mapped set of governing equations is expressed as
\begin{align}
    \frac{\partial\tilde{\vector{U}}}{\partial t}
    +\frac{\partial\tilde{\vector{F}}^{a}}{\partial \xi}
    +\frac{\partial\tilde{\vector{G}}^{a}}{\partial \eta}
    +\frac{\partial\tilde{\vector{H}}^{a}}{\partial \zeta}
    =
     \frac{\partial\tilde{\vector{F}}^{d}}{\partial \xi}
    +\frac{\partial\tilde{\vector{G}}^{d}}{\partial \eta}
    +\frac{\partial\tilde{\vector{H}}^{d}}{\partial \zeta}
    +\tilde{\vector{S}},
\end{align}
where
\begin{align}
    &\tilde{\vector{U}}=\frac{\vector{U}}{J},\label{eqn:U}\\
    &\tilde{\vector{F}}^{a,d}=\frac{\xi_x}{J}{\vector{F}}^{a,d}
    +\frac{\xi_y}{J}{\vector{G}}^{a,d}+\frac{\xi_z}{J}{\vector{H}}^{a,d},\\
    &\tilde{\vector{G}}^{a,d}=\frac{\eta_x}{J}{\vector{F}}^{a,d}
    +\frac{\eta_y}{J}{\vector{G}}^{a,d}+\frac{\eta_z}{J}{\vector{H}}^{a,d},\\
    &\tilde{\vector{H}}^{a,d}=\frac{\zeta_x}{J}{\vector{F}}^{a,d}
    +\frac{\zeta_y}{J}{\vector{G}}^{a,d}+\frac{\zeta_z}{J}{\vector{H}}^{a,d},\\
    &\tilde{\vector{S}}=\frac{\vector{S}}{J}.
\end{align}
$\tensor{J}$ is the Jacobian of the transformation.
\begin{equation}
\tensor{J}
=
\begin{bmatrix}
\xi_x & \xi_y & \xi_z\\
\eta_x & \eta_y & \eta_z\\
\zeta_x & \zeta_y & \zeta_z
\end{bmatrix}.
\end{equation}
We denote $|\tensor{J}|=J$.
After this derivation, the standard conservative FD method for Cartesian grids used in \citep{DiRenzo20} can be adapted to discretize the spatial derivatives of the mapped fluxes.
The inviscid fluxes are computed using WENO3-Z/TENO6.
A relevant assessment on WENO3-Z and TENO6 is reported in \citep{Maeda22ctr}.
The diffusive fluxes are obtained by a standard second-order central-difference scheme.
The variables on the original domain can be readily obtained by the inverse transformation.
In the simulations presented in this study, the temporal integration of the equations is realized by the third-order strong-stability-preserving Runge–Kutta (SSP-RK3) method \citep{Gottlieb01}, while other integration methods can be readily used. Specifics of implementations are available upon request.

\subsection{Solver Performance Analysis}
\subsubsection{Accuracy}
Periodic domains are used.
To generate the skewed grids used in the convergence tests, we map a $N\times N$ square uniform grid defined on a domain, $x,y\in[-L/2,L/2]$, by the following transformation \citep{ham04}:
\begin{align}
    \xi(x,y) = x[1 + \beta\mathrm{sin}(2\pi y/L)],\hspace{1em}\eta(x,y) = y[1 + \beta\mathrm{sin}(2\pi x/L)],\label{eqn:grid}
\end{align}
where $L$ is the width of the domain.
$\beta=0.2$ is used.
The numerical solutions are compared to the analytical solution, by increasing the resolution.
For the vortex advection problem, the initial condition is defined as
\begin{equation}
[u,v,T]=[\frac{\beta}{2\pi}e^{\alpha[1-(x^2+y^2)/r_0^2]}y+u_0,-\frac{\beta}{2\pi}e^{\alpha[1-(x^2+y^2)/r_0^2]}x+v_0,-\frac{(\gamma-1)\beta^2}{16\alpha\gamma\pi^2}e^{2\alpha[1-(x^2+y^2)/r_0^2]}+T_0],
\end{equation}
where $\alpha$ defines the vortex decay rate and $\beta$ is the vortex strength. $r_0$ is the radius of the vortex core. The pressure and the density may be obtained from the ideal gas law $p=\rho T$ and the isentropic relationship $p=\rho^\gamma$. $r_0=0.5$ is chosen to initially confine the vortex in the central region of the domain where the mesh is highly skewed.
We choose $\beta=5.0$, $\alpha=1/2$, and $\gamma=1.4$.
The domain size is defined by $L=10$.
The reference values are chosen as $[u_0,v_0,T_0]=[1.0,1.0,1.0]$.
The error is obtained at $t=0.05$ unit time. A sufficiently small, constant time-step of $\Delta t =2.0\times10^{-5}$ is used for all $N$.

For the two-dimensional Taylor-Green vortex problem, the initial condition is defined as
\begin{equation}
[u,v,p]=[u_0\mathrm{sin}\left(\frac{2\pi x}{L}\right)\mathrm{cos}\left(\frac{2\pi y}{L}\right),-u_0\mathrm{cos}\left(\frac{2\pi x}{ L}\right)\mathrm{sin}\left(\frac{2\pi  y}{L}\right),p_0-\frac{\rho_0u_0^2}{4}\left(\mathrm{cos}\frac{4\pi  x}{L}\right)\mathrm{cos}\left(\frac{4\pi y}{L}\right)],
\end{equation}
where $u_0$, $p_0$, $\rho_0$ are the reference velocity, pressure, and density, respectively. The reference values are chosen such that the vortex is evolved with a low reference Reynolds number of ${Re}_0(=\rho_0u_0L/\mu)=L$ and an effectively incompressible Mach number of ${Ma}_0(=u_0/\sqrt{\gamma p_0/\rho_0})=10^{-2}/\sqrt{\gamma}$, with the reference viscosity of $\mu=1.0$ and $\gamma=1.4$.
The domain size is defined by $L=2\pi$.
The error is obtained at $t=0.05$ unit time. A sufficiently small, constant time-step of $\Delta t=1.0\times10^{-5}$ is used for all $N$.

\subsubsection{Weak scaling analysis}
The weak scaling analysis was performed on Lassen, a hybrid GPU-CPU machine at Lawrence Livermore National Laboratory (LLNL). The machine is equipped with four Nvidia Tesla V100 GPUs per node. We simulate a single-component gas flow with increasing numbers of GPUs up to 512, and with $216^3$ grids per GPU, and obtain the execution time per iteration.

\subsection{Diffusion flame}
\subsubsection{Physical model and numerical setup}
To model the laser-deposition in the energy source term, we employ the following expression for the heat source term convoluted with a Gaussian kernel in both space and time, which respectively model the size of the laser's focal area and the temporal spread (duration) of the energy deposition.
\begin{equation}
q_L=\frac{E_{L}}{(2\pi)^{3/2}\sigma_r^2\sigma_t}e^{-\frac{1}{2}\left(\frac{r}{\sigma_r}\right)^2}e^{-\frac{1}{2}\left(\frac{t-t_0}{\sigma_t}\right)^2},
\label{eq:laser-qs}
\end{equation}
where $E_{L}$ is the amount of energy deposited per unit depth of the domain, $\sigma_r$ and $\sigma_t$ are the spatial and temporal support of the kernel. $r$ is the spatial distance from the center of the ignition kernel.
Similar approaches have previously been taken for modeling the laser-induced ignition \citep{Lacaze09}.

A jet of pure gaseous \ce{O2} is horizontally injected in the domain filled with quiescent, gaseous \ce{CH4} from the center of the left boundary.
The co-flow of gaseous \ce{CH4} is injected from the rest of the left domain boundary with a velocity much smaller than the jet.
The left and the right domain boundaries are modeled using the Navier-Stokes characteristic boundary condition (NSCBC) \citep{Poinsot92,okongo2002}.
The other boundaries are modeled as no-slip walls.
A similar setup of realistic diffusion flame can be found, for example, in \citep{Pantano04}.
The stability of a similar flame was studied, for instance, by \citep{Furi02,See14}.

TENO6 is employed to compute the inviscid fluxes.
The one-step global mechanism is employed to model the chemistry.
The energy deposition is modeled by an intense, localized, short-time energy deposition in the domain.
The domain size, resolution, the jet and co-flow velocities, and the location, timing, and duration of the energy deposition can be specified by the users in the input file, along with other parameters.
An input file for this case is provided in supplementary information (S.2.1).

\subsubsection{Data management}
For the bisection line search that is employed to identify the minimum ignition amplitudes, we tightens an interval $[a,b]$ such that laser energy $a$ does not ignite while $b$ ignites. This requires far fewer simulations to accurately estimate the minimum laser energy for ignition. To manage these simulations, we use the Fireworks ensemble management framework~\cite{fireworks}.

\subsubsection{Autoencoder}
The autoencoder, inspited by \citep{Lee20}, consists of six convolutional layers, each halving the physical dimensions while keeping 32 channels of information.
After these layers, the result is fed through a fully connected network that results in a $d$-dimensional code (latent variables).
This network topology is repeated in reverse to decode the code into a full representation of the state.
This network is then trained to minimize the least squares mismatch
between the input to the network (the state) and the output of the network (the approximated state).

\subsection{Model rocket combustor}

\subsubsection{Physical model}
The combustor is initialized at 300\,K with pure \ce{O2} and zero velocity everywhere, and injection of \ce{CH4} and \ce{O2} begins at $t=0$.  The initial ambient pressure 137.5\,kPa corresponds to the pre-ignition pressure measured in the companion experiment (Fig.~\ref{fig4}a).  The mass flow rate and temperature of each stream --- 6.44\,g/s and 242\,K for \ce{O2}, and 1.66\,g/s and 282\,K for \ce{CH4} --- are kept constant throughout the injection and ignition stages.  The initial inflow velocities are 295\,m/s for \ce{O2} and 191\,m/s for \ce{CH4}, corresponding to Mach 1.0 and 0.44 respectively, are decreased in order to maintain a constant mass flow rate as the chamber pressure increases.  The \ce{O2} jet has diameter 3.57\,mm, and the annular \ce{CH4} co-flow has inner and outer diameters 5.33\,mm and and 6.35\,mm, respectively.  The jet Reynolds number based on the initial \ce{O2} velocity and jet diameter is 66,000.  The combustor and exit nozzle have an cross-sectional area ratio of 64:1.

The injection stage proceeds for 2.8\,ms, during which the concentration of \ce{CH4} in the combustor increases and large recirculation zones are established.  Without any external energy addition during this stage, chemical reactions are negligible, and the flow is modeled as a chemically inert mixture of \ce{CH4} and \ce{O2}.
The laser is deployed after this injection stage at $t_L$, and the subsequent ignition dynamics are modeled with the one-step combustion model.
The laser-energy deposition is modeled as a localized energy source,
\begin{equation}
   \dot{q}_L = \dot{e}_Lf(\vector{x})e^{-\frac{1}{2}\left(\frac{t-t_L}{\sigma_t}\right)^2}
    \label{eq:laser-Qdot}
\end{equation}

\noindent
where $\sigma_t=4.2$\,ns is the laser pulse duration, $f(\vector{x})\in[0,1]$ determines the geometry of the energy kernel, and $\dot{e}_L$ is the volumetric rate of energy deposition, a parameter used to control the total energy deposited $E_L=\int\int \dot{q}_L \mathrm{d}V\,\mathrm{d}t$.  The size of the energy kernel is varied from 1.0\,mm to 2.4\,mm in E-1, with larger kernels corresponding to greater $E_L$, and fixed at 1.8\,mm for E-2.  The laser pulse duration, kernel size, and energy are consistent with those of the companion experiment.  
This model differs critically from \eqref{eq:laser-qs} in the prescription of $f(\vector{x})$: asymmetry in the geometry of the energy kernel can lead to laser-generated flow that transports hot gas over distances much larger than the initial kernel and impacts the ignition outcome \citep{morsy2002,bradley2004,Phuoc06}.  Additional details on the laser model, the simulation setup, and ignition dynamics can be found elsewhere  \citep{wang2020jfm,wang2021cnf,wang2022ctr}.

\subsubsection{Numerical setup}
The combustor is discretized with a curvilinear mesh whose cross-section is rounded to approximate the internal geometry of the corresponding experimental combustor (Fig.~\ref{fig4}).  The mesh consists of 98 million points (${960\times320\times320}$); the finest mesh spacing is $\Delta x=\Delta y=\Delta z=88\,\mu\text{m}$ at the injector orifice, which corresponds to 12 points across the annular thickness of the \ce{CH4} co-flow and 41 points across the \ce{O2} jet.  Away from the near-injector region, resolution requirements are less severe, and the mesh is stretched.

Inviscid fluxes are computed using WENO3-Z.  The injector inlet and nozzle outlet, corresponding to the left and right boundaries of the simulation domain, are modeled using the NSCBC approach for reacting flows \citep{okongo2002}, and all walls are no-slip and isothermal with temperature 300\,K.

\subsubsection{Ensemble simulation}
In all cases designated as ignition failure, the failure was confirmed by the global maximum temperature decreasing below 600\,K, which is much lower than the temperature necessary for ignition.
Each simulation uses 48 Tesla V100 GPUs on Lassen at LLNL.  A single simulation of the pre-ignition, injection stage ($t\leq 2.8\,\text{ms}$) was conducted, which required 50 hours of run time. Each simulation of the post-ignition stage requires up to 10 hours, depending on the ignition delay. In total, $O(10^4)$ GPU-hours were used for 24 simulations of the 3D combustor.

\section*{Acknowledgement}
This investigation was funded by the Advanced Simulation and Computing (ASC) program of the US Department of Energy’s National Nuclear Security Administration (NNSA) via the PSAAP-III Center at Stanford, Grant No. DE-NA0002373.
The authors thank the members of the PSAAP-III Center for numerous fruitful discussions.

\bibliographystyle{abbrv}

\end{document}


\begin{frontmatter}


\title{Supplementary information: An integrated heterogeneous computing framework for ensemble simulations of laser-induced ignition}

\author[1]{Kazuki Maeda\corref{cor1}}
\ead{kemaeda@stanford.edu}
\cortext[cor1]{Corresponding author}
\author[2]{Thiago Teixeira}
\ead{thiagoxt@stanford.edu}
\author[1]{Jonathan M. Wang}
\ead{jmwang14@stanford.edu}
\author[3]{Jeffrey M. Hokanson}
\ead{jeho8774@colorado.edu }
\author[4]{Caetano Melone}
\ead{cmelone@stanford.edu}
\author[5]{Mario {Di Renzo}}
\ead{direnzo.mario1@gmail.com }
\author[6]{Steve Jones}
\ead{stevejones@stanford.edu }
\author[1]{Javier Urzay}
\ead{jurzay@stanford.edu }
\author[6]{Gianluca Iaccarino}
\ead{jops@stanford.edu}

\address[1]{Center for Turbulence Research, Stanford University, USA}
\address[2]{Department of Computer Science, Stanford University, USA}
\address[3]{Smead Aerospace Engineering Sciences, University of Colorado at Boulder, USA}
\address[4]{High Performance Computing Center, Stanford University, USA}
\address[5]{Centre Européen de Recherche et de Formation Avancée en Calcul Scientifique, France}
\address[6]{Department of Mechanical Engineering, Stanford University, Stanford University, USA}

\end{frontmatter}

\tableofcontents

\setcounter{section}{0}
\setcounter{figure}{0}
\renewcommand{\thesection}{S.\arabic{section}}
\renewcommand{\figurename}{Figure S}

\clearpage
\newpage
\section{Performance verification and analysis}
\subsection{Weak scaling analysis}
\begin{figure}[h]
    \centering
    \subfloat{\includegraphics[width=70mm,trim=0 00mm 0 0mm, clip]{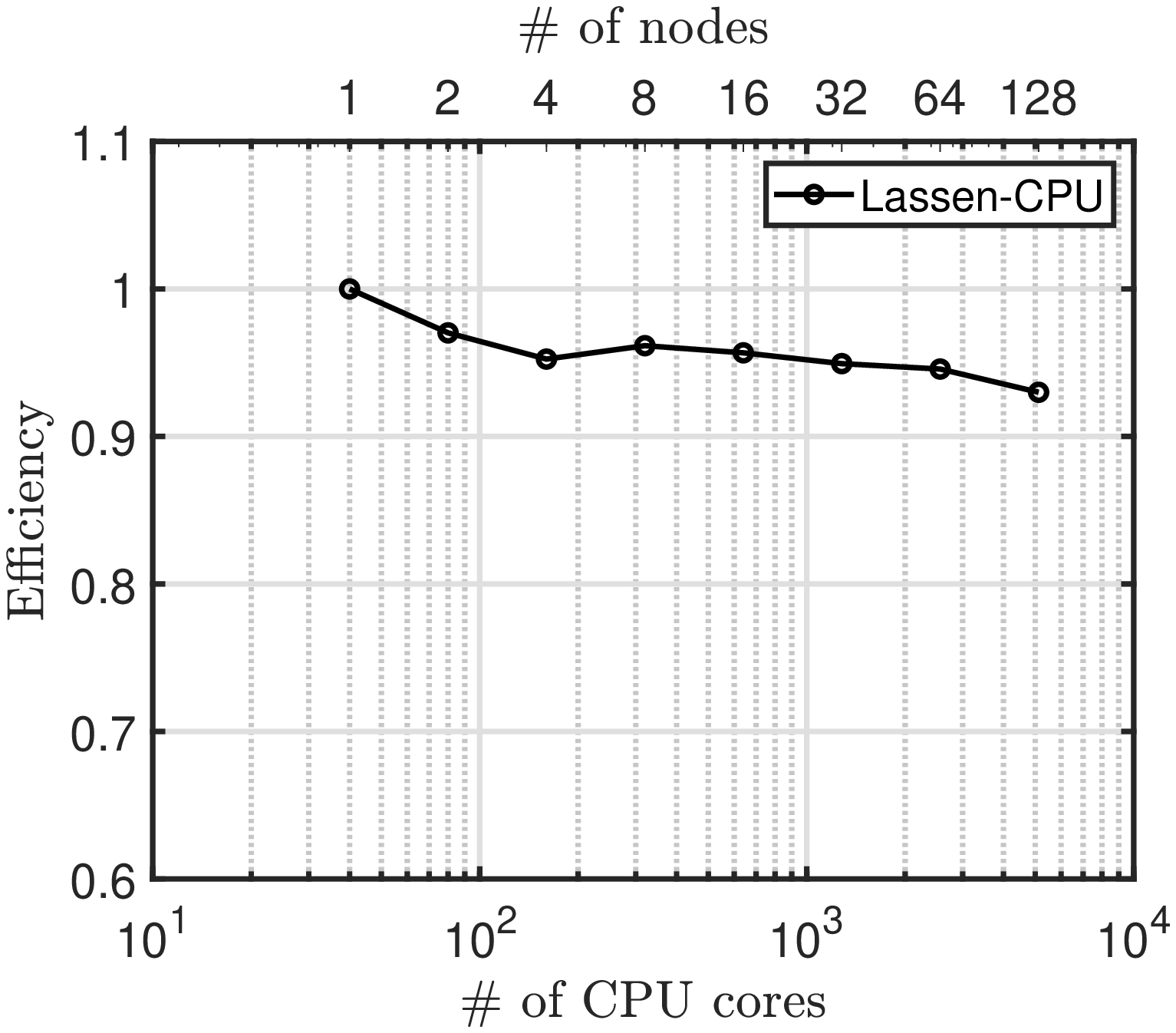}}\\
    \subfloat{\includegraphics[width=70mm,trim=0 00mm 0 0mm, clip]{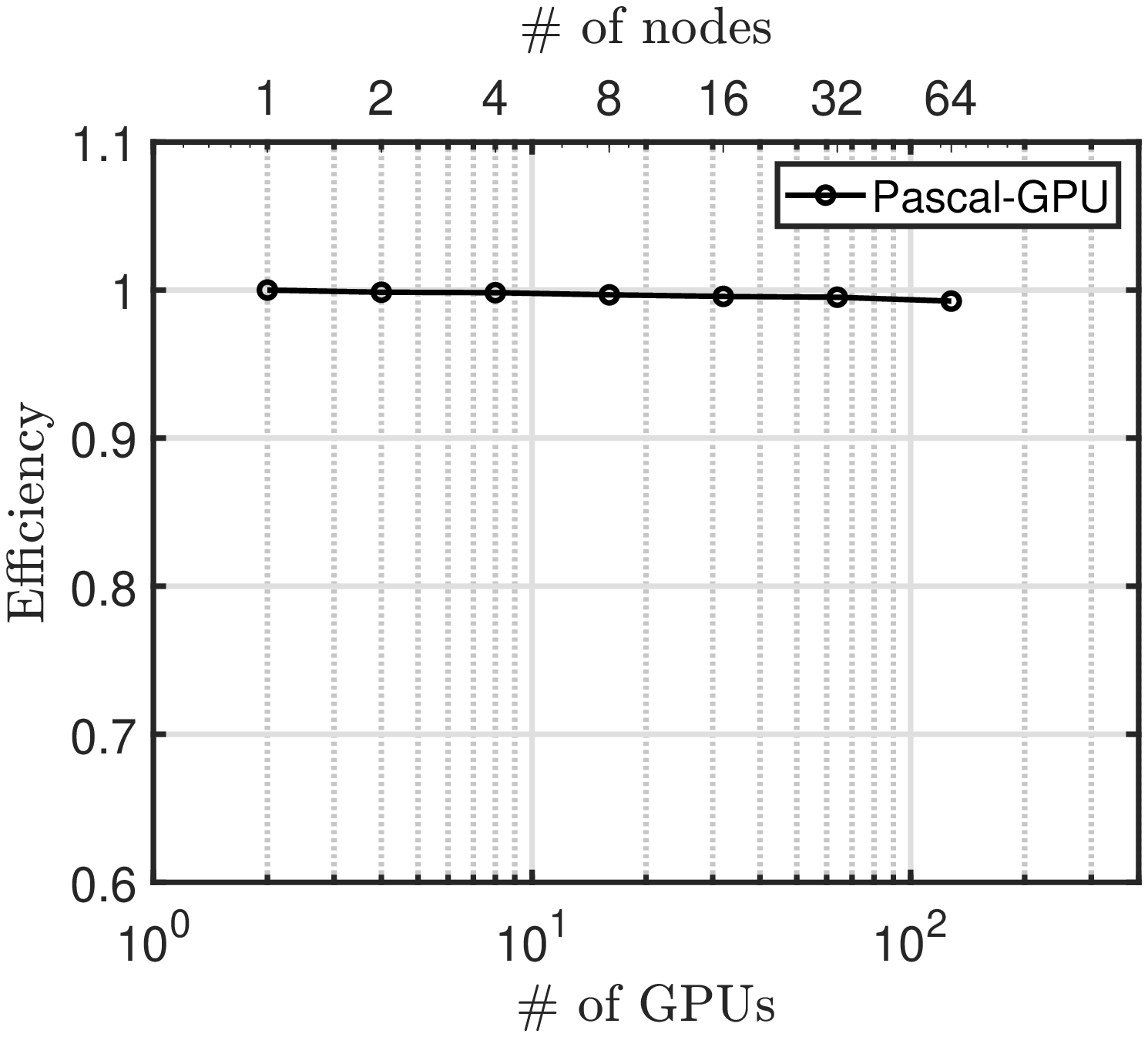}}
    \subfloat{\includegraphics[width=70mm,trim=0 00mm 0 0mm, clip]{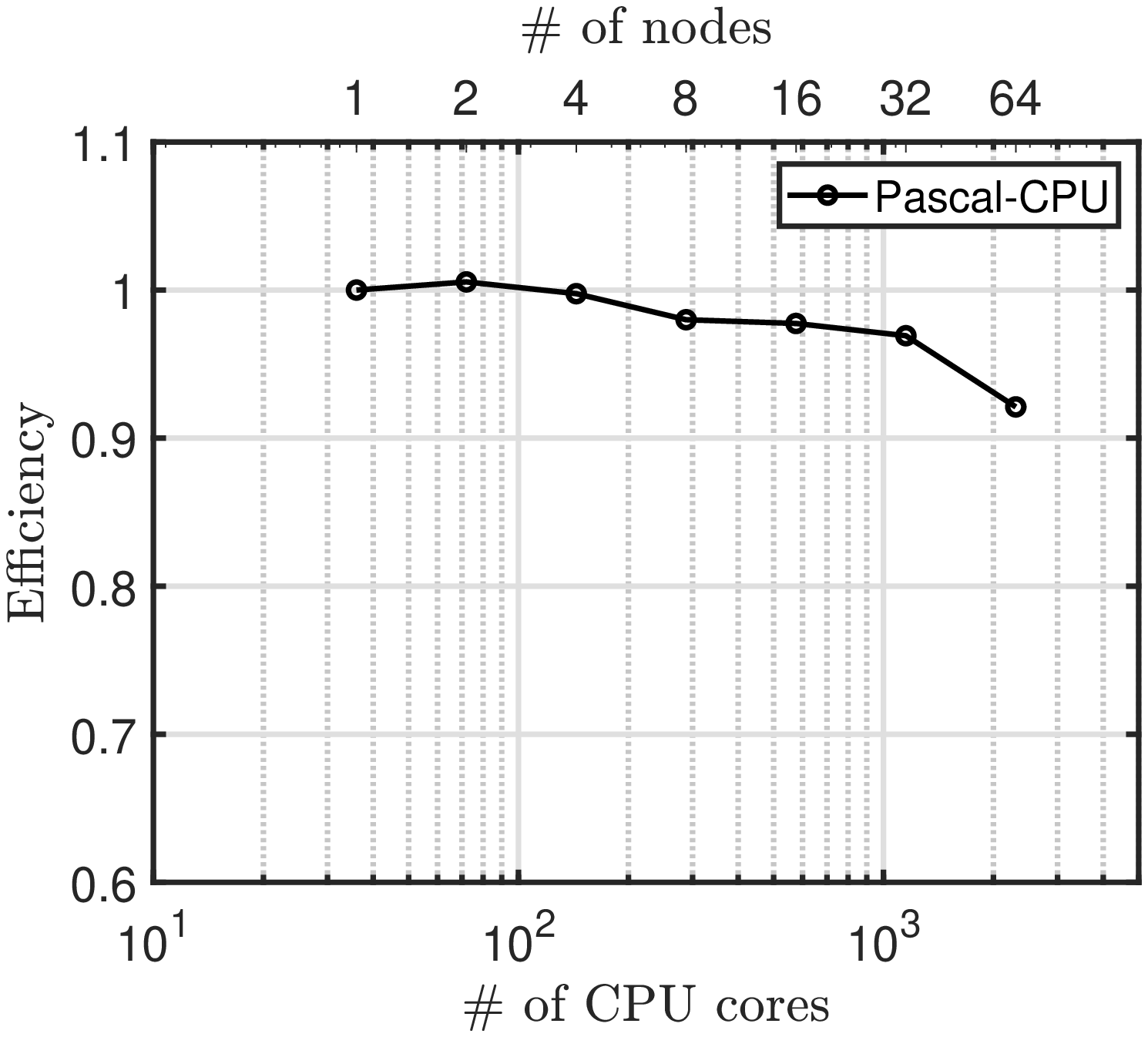}}
    \caption{Supplementary weak scaling data on Lassen and Pascal at LLNL.}
    \label{fig:cpu}
\end{figure}
Figure S\ref{fig:cpu} shows the weak scaling plots of HTR using (a) CPUs on Lassen (Two IBM Power9 Cores/Node, totalling 40 cores/Node), (b) GPUs on Pascal (two Nvidia Tesla P100 GPUs/Node), and (c) CPUs on Pascal (two Intel Xeon E5-2695 v4 CPUs/Node, totalling 36 cores/Node). Pascal is a CPU/GPU machine at LLNL. Detailed specification of Pascal is available elsewhere\footnote{https://hpc.llnl.gov/hardware/compute-platforms/pascal (accessed: 2021-12-31)}. The efficiency unit is defined based on the simulation using a single node, following Fig. 2d. The test problem used in these analyses follows section 3.2.1. For Lassen-CPU, $36^3$ grids/core are used. For Pascal-GPU, $216^3$ grids/GPU are used. For Pascal-CPU, $64^3$ grids/cores are used.
An excellent scaling is obtained on Pascal-GPU up to 128 GPUs (64 nodes), similar to the result of Lassen-GPU (Fig. 2d). Note that 163 GPUs are available on Pascal. Although the the scaling of CPUs are not as efficient as GPU counterparts on both machines, the efficiency is higher than 90\% up to 5120 CPU cores (128 nodes) on Lassen and up to 2304 CPU cores (64 nodes) on Pascal. Overall, results indicate favorable weak scaling on multiple CPU/GPU machines.

\clearpage
\subsection{Unit testing on multiple machine environments}
\begin{figure}[h]
    \centering
    \includegraphics[width=80mm,trim=0 00mm 0 0mm, clip]{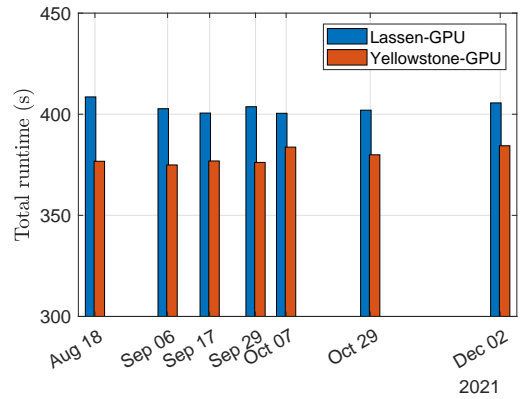}
    \caption{Progression of the run-time required for unit tests on Lassen and Yellowstone. GPUs are used on both machines.}
    \label{fig:test}
\end{figure}
Figure \ref{fig:test} shows the progression of the run-time required for HTR's unit test in Aug-Dec 2021 on Lassen and Yellowstone. Data on representative dates are shown. The same configuration is used for all data. An example configuration of the unit test can be found in the open-source version of the solver \footnote{https://github.com/stanfordhpccenter/HTR-solver/tree/master/unitTests (accessed: 2021-12-31)}. Yellowstone is a local cluster at Stanford University. Nvidia Tesla V100 GPUs are used on Lassen. Nvidia Titan X Maxwell GPUs are used on Yellowstone. For each test date, the latest commit of Legion's control\_replication branch\footnote{https://gitlab.com/StanfordLegion/legion/-/tree/control\_replication (accessed: 2021-12-31)} is used as of the date.
The fluctuation of the run-time can be explained by updates of Legion, dependencies, and machine environments.
The magnitude of the fluctuations relative to the total time is nominally small, indicating the small influence of the changes.

\clearpage
\newpage
\subsection{Legion Spy: Task graph}
\begin{figure}[h]
    \centering
    \includegraphics[width=164mm,trim=0 0mm 0 0mm, clip]{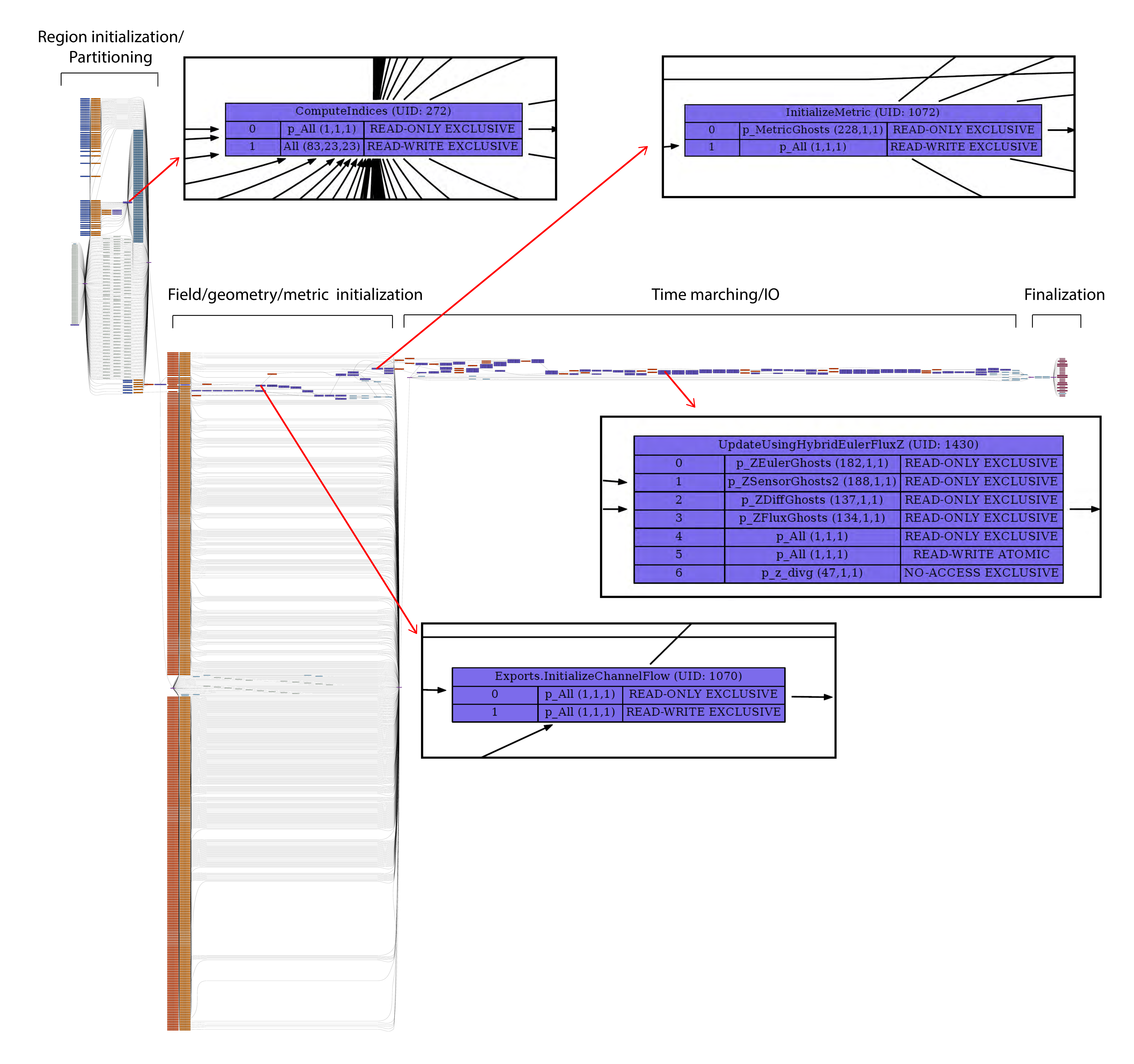}
    \caption{Task graph of a representative simulation of HTR obtained using the Legion Spy tool.}
    \label{fig:graph}
\end{figure}
Figure \ref{fig:graph} shows the task graph of a representative simulation of HTR obtained using the Legion Spy tool\footnote{https://legion.stanford.edu/debugging/\#legion-spy (accessed: 2021-12-31)}.
The sequence of task trees are displayed during the run-time from left to right.
Data partitions used by each task are shown with access privileges.

\clearpage
\newpage
\subsection{Legion Prof: Profiling of CPU/GPU co-processing}
\begin{figure}[h]
    \centering
    \includegraphics[width=164mm,trim=30 120mm 0 80mm, clip]{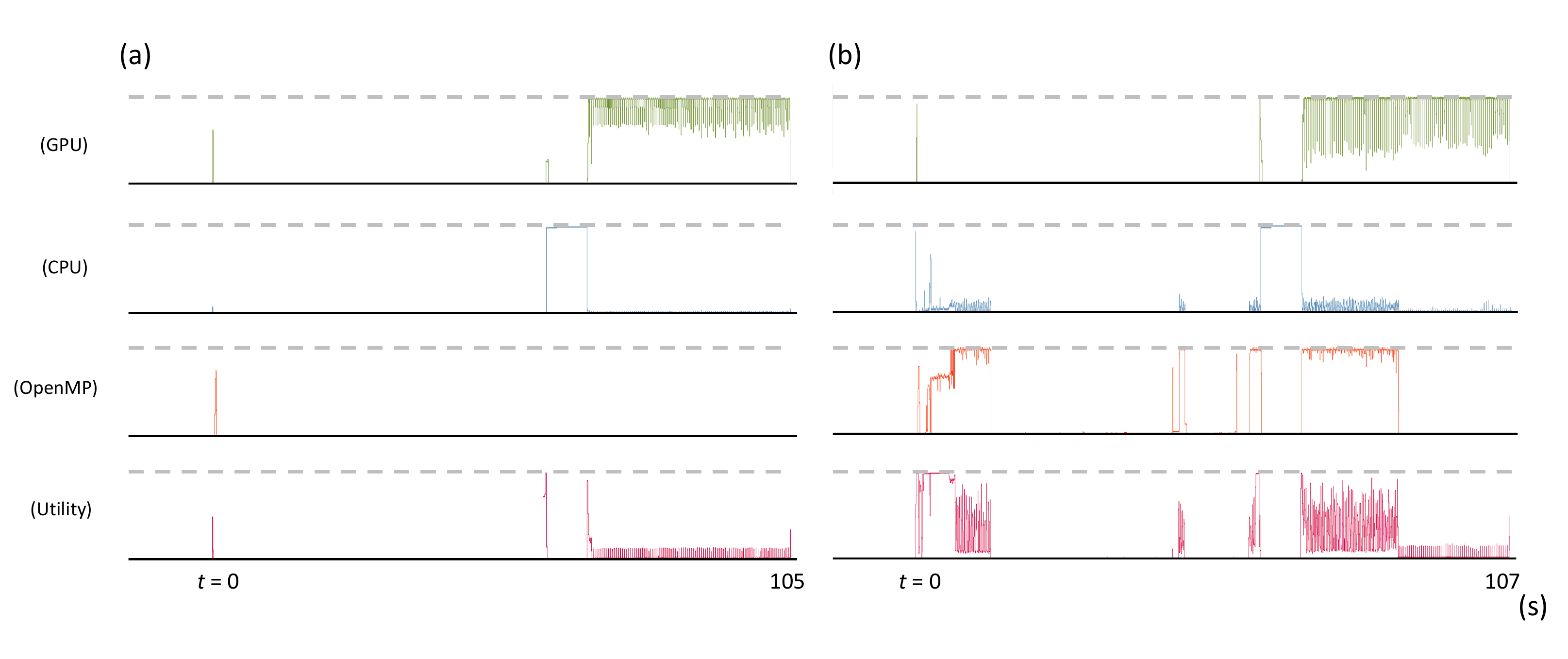}
    \caption{Results of run-time profiling of ensembles with (a) no LF sample and b) 8 LF samples, from the runs shown in Fig. 2e. The utilization of CPU, CPU, OpenMP, and Utility (CPUs allocated for task handling) are shown. Dotted lines indicate 100\% usage.}
    \label{fig:prof}
\end{figure}
Figure S\ref{fig:prof} shows the run-time profiling of the ensembles shown in Fig. 2e with (Fig. S\ref{fig:prof}a) no LF sample and (Fig. S\ref{fig:prof}b) 8 LF samples, respectively, obtained using the Legion Prof tool \citep{Legion}\footnote{https://legion.stanford.edu/profiling/index.html (accessed: Dec 31, 2021)}.
The usage of a representative GPU, CPU, OpenMP processor, and Utility is shown for both samples.
In Legion, OpenMP processors are defined on CPUs and are regarded as independent processing units defined for many run-time tasks.
Utility occupies a small portion of the CPU resource and is used for the task scheduling on processing units.
The usages of the GPU and that of the CPU are nearly identical for both cases. The GPU usage is first nearly zero due to initialization, but becomes high later.
Similarly, for both cases, the CPU is active right before the active period of the GPU, and nearly zero elsewhere.
We see clear differences in the usages of OpenMP and Utility between the ensembles.
Without the LF sample (Fig. S\ref{fig:prof}a), the usage of OpenMP is nearly zero. That of Utility is uniformly small during the activation of GPU and nearly zero otherwise.
With LF samples (Fig. S\ref{fig:prof}b), however, OpenMP is highly active at the initial period and also during the period of GPU activation. Utility follows a similar trend. This activation indicates the additional task loading on CPUs (most of the task executions occur on OpenMP processors) and that the tasks on CPU/OpenMP do not affect the total runtime.

\clearpage
\newpage
\section{Diffusion flame case}
\subsection{Input file for the diffusion flame case}
Below we show the entries of the input file for the diffusion flame case.
\begin{minted}[fontsize=\scriptsize,bgcolor=lightgray]{python}
  1 {
  2  "Case" : {
  3    "ReInlet"     : 400,    % Reynolds number of the O2 jet
  4    "Ma_F"        : 0.001,  % Mach number of the CH4 jet
  5    "Ma_Ox"       : 0.1,    % Mach number of the O2 jet
  6    "TInf"        : 350.0,  % Background temperature
  7    "PInf"        : 50662.5 % Background pressure
  8    },
  9
 10    "Mapping" : {
 11       "wallTime" : 600 % Maximum simulation wall clock time (minutes)
 12    },
 13
 14    "Grid" : {
 15       "xNum" : 500,    % Resolution in the x-direction
 16       "yNum" : 250,    % Resolution in the y-direction
 17       "xWidth" : 32.0, % Width of the domain in the x-directon (unit: jet thickness)
 18       "yWidth" : 16.0  %  Width of the domain in the y-direction (unit: jet thickness)
 19    },
 20
 21    "Integrator" : {
 22       "startIter" : 0,   % Time step at the simulation start
 23       "startTime" : 0.0  % Simulation start time
 24    },
 25
 26    "Flow" : {
 27       "laserIgnition" : {
 28           "Amplitude" : 0.05,       % Laser intensity
 29           "Center" : [3.0,0.5,0.0], % Coordinate of the deposition spot [x,y,z(=0)] (unit: jet thickness)
 30           "Radius" : 0.25,          % Spatial radius of the kernel (unit: jet thickness)
 31           "Delay" : 9.6,            % Delay of the energy deposition after t=0 (unit: [jet thickness/jet velocity])
 32           "Duration" : 0.5          % Duration of the deposition (unit: jet flow through time)
 33       }
 34    },
 35
 36    "IO" : {
 37       "wrtRestart" : true,             % Write restart file
 38       "restartEveryTimeSteps" : 100,   % Data output interval
 39       "probesSamplingInterval" : 2,    % Probe sampling interval
 40       "probes" : [{"fromCell":[0,0,-2], "uptoCell":[1,1,2]}] % Probe coordinate
 41    }
 42 }
\end{minted}
\normalsize

\clearpage
\newpage
\subsection{Autoencoder}
\begin{figure}[h]
    \centering
    \includegraphics[width=70mm,trim=0 0mm 0 0mm, clip]{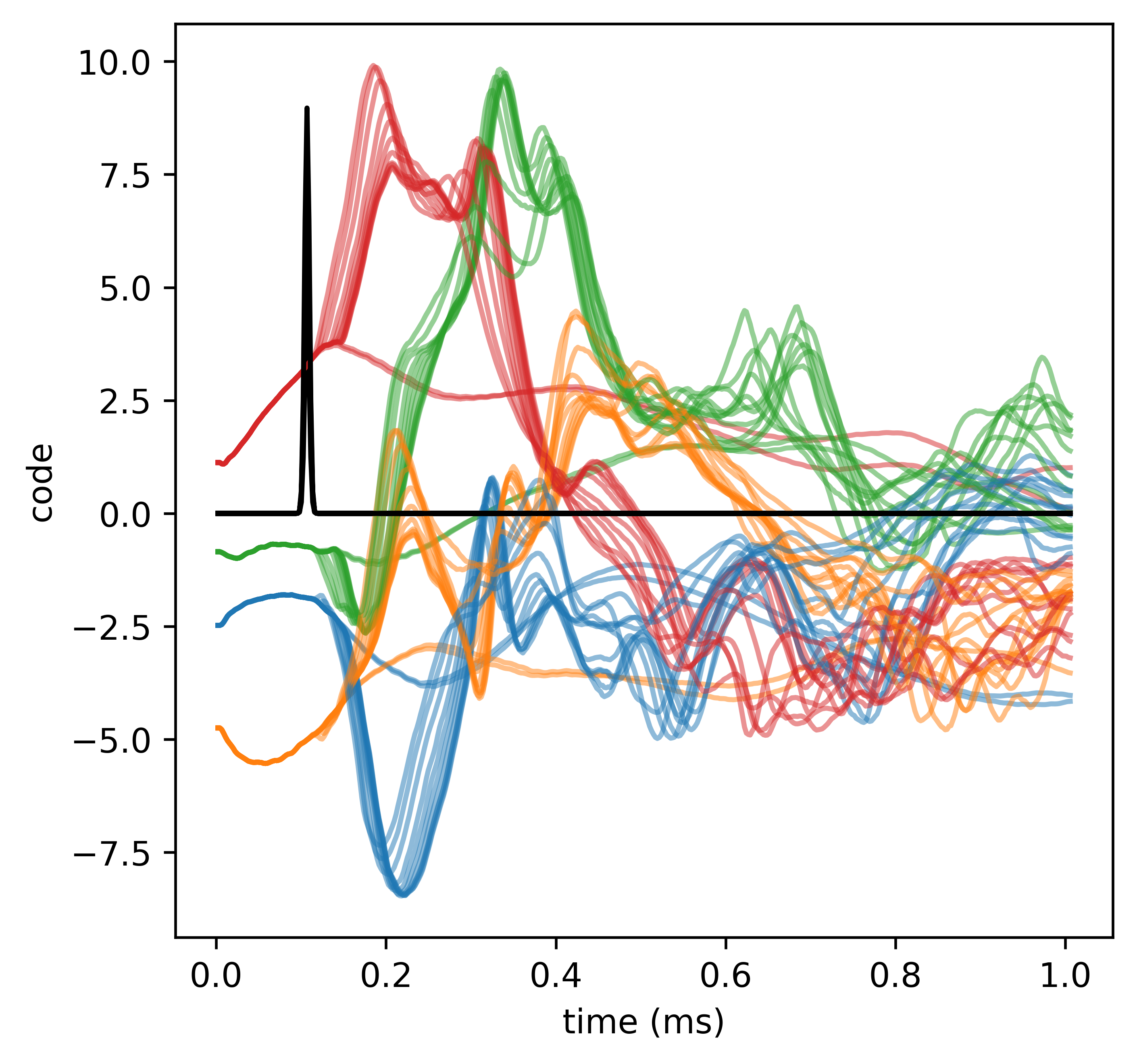}
    \caption{Evolution of representative codes (latent variables) of the autoencoder learned from the data of the diffusion flame case.}
    \label{fig:code}
\end{figure}
Figure S\ref{fig:code} shows the evolution of representative codes (latent variables) of the autoencoder for the simulations with various laser amplitude, for the diffusion flame case The peak of the black line indicates the timing of the laser deployment. The parameters vary smoothly in time and with varying laser energy.

\clearpage
\newpage
\section{Combustor simulations}
\subsection{Setup}
\begin{figure}[h]
    \centering
    \includegraphics[width=0.8\textwidth]{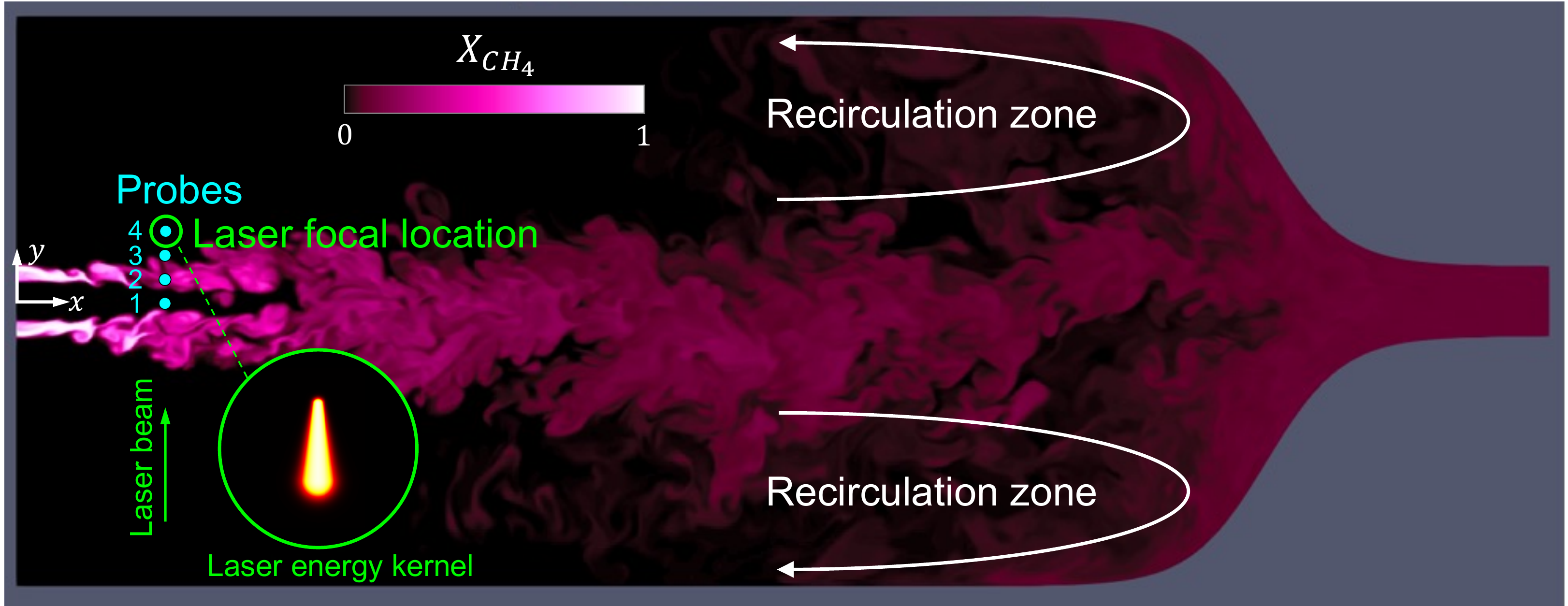}
    \caption{\ce{CH4} mole fraction in the 3D combustor simulation at $z=0$ and $t=2841\,\mu\text{s}$.  The probes are located at $x=12.7\,\text{mm}$, or two jet diameters downstream of the injector, and are evenly spaced between the centerline and laser focal location at $y=6\,\text{mm}$.  The inset shows the laser-energy kernel.}
    \label{fig:probe-locations}
\end{figure}

The laser-energy kernel has the prescribed geometry shown in figure S\ref{fig:probe-locations}.  The asymmetry in the direction of the laser beam is characteristic of laser-induced breakdown at these conditions, and the model has been shown to reproduce key laser-generated flow features in both non-reacting \citep{wang2020jfm} and reacting \citep{wang2021cnf} settings.  Although the laser is focused outside the \ce{CH4}--\ce{O2} shear layer (Fig. S\ref{fig:probe-locations}), it generates flow that transports hot gas toward the fuel and ignites it; details of these dynamics can be found in a separate investigation of the combustor ignition \citep{wang2022ctr}.  The time traces of local flow quantities are measured at the four probe locations shown in Fig. S\ref{fig:probe-locations}, which are used to determine $t_L$ of the second ensemble.

\clearpage
\newpage
\subsection{Probe signals}
\begin{figure}[h]
    \centering
    \includegraphics[width=\textwidth]{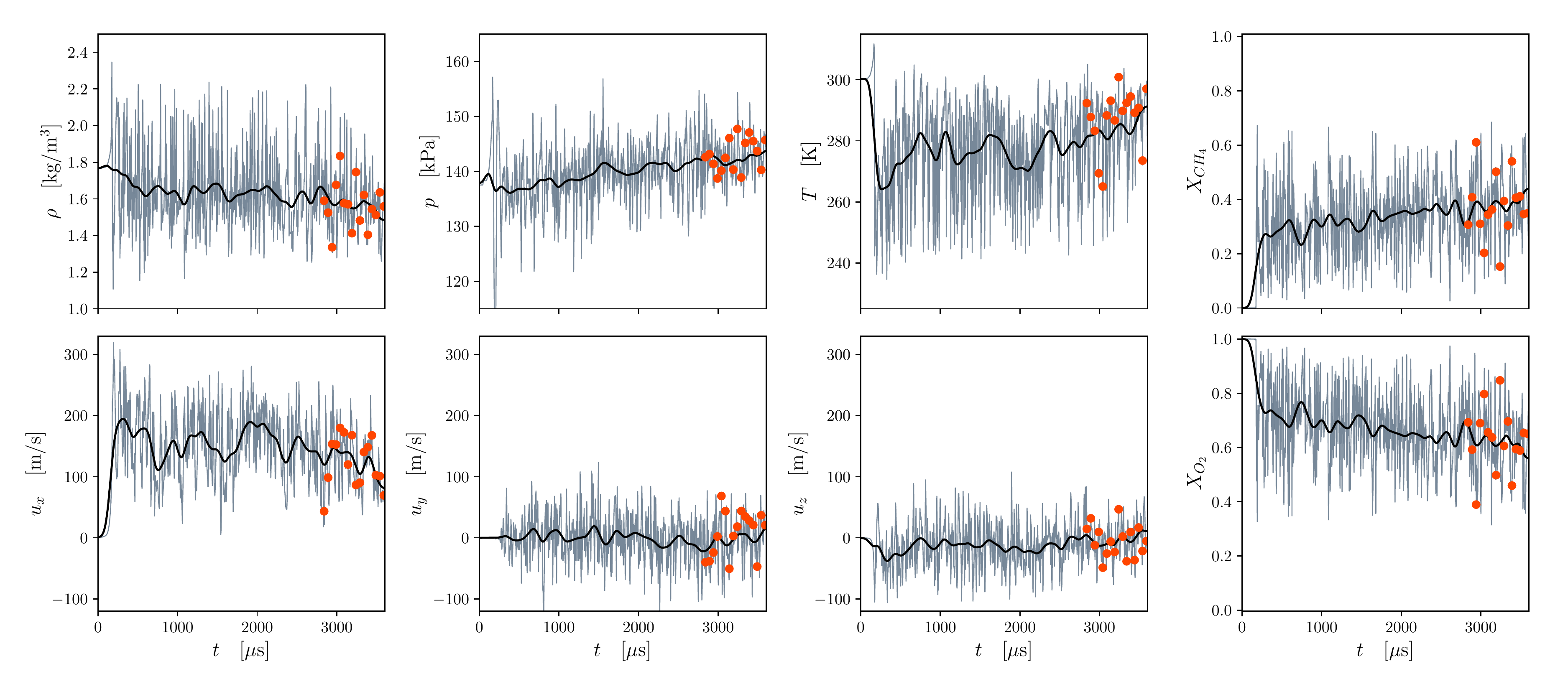}
    \caption{Time trace at probe 1 in Fig. S\ref{fig:probe-locations}.  Markers indicate $t_L$ for the 16 simulations of the second ensemble, and the thick solid line is a moving average with a time window of $50\,\mu\text{s}$.}
    \label{fig:probe0}
\end{figure}

Figure S\ref{fig:probe0} shows the raw data obtained from the probe on the centerline (probe 1 in Fig. S\ref{fig:probe-locations}) as well as moving average with a window size of 50 $\mu$s.
For all quantities, we clearly observe large-amplitude, high-frequency fluctuations in the raw data and the transient, non-stationary growth in the moving average plots.
The timings of laser deployment in the samples in the second ensemble, which are indicated by the circular dots, are selected late in the time history ($t\geq 2800\,\mu\text{s}$) in order to allow time for large-scale recirculation zones to be established (Fig.~S\ref{fig:probe-locations}), as these flow structures are anticipated to play a role in the growth of the flame as it fills the combustor.

\clearpage
\newpage
\begin{figure}[h]
    \centering
    \includegraphics[width=\textwidth]{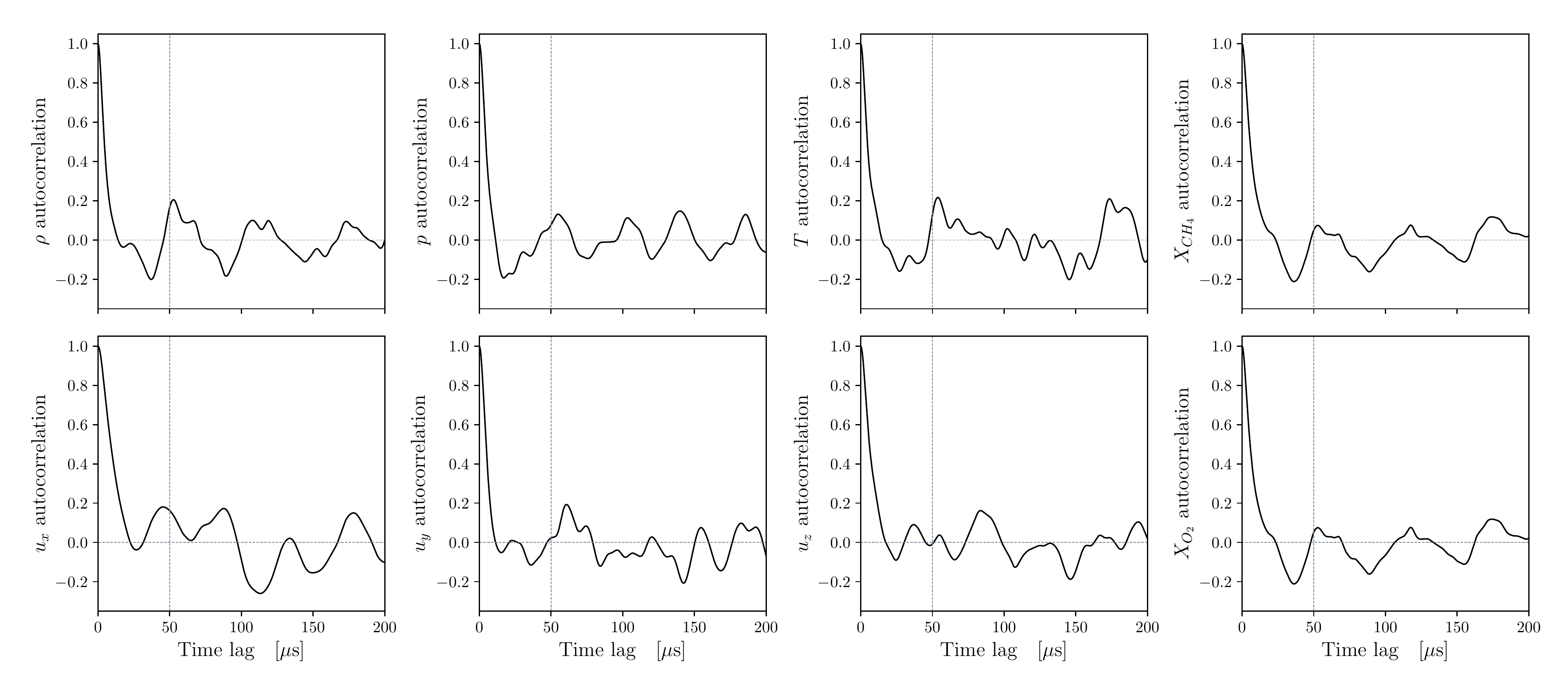}
    \caption{Auto-correlation of the measurements at probe 1 within a window of $t:t\in[2800,3600]$ $\mu$s.}
    \label{fig:autocorr0}
\end{figure}
Figure S\ref{fig:autocorr0} shows the auto-correlations \citep{Bracewell86} of the probe data (shown in Fig. S\ref{fig:probe0}) within the window of $t:t\in[2800,3600]$ $\mu$s, in which we assume that the turbulence-induced fluctuations are statistically stationary.
All flow quantities show relatively low auto-correlation for intervals greater than 50 $\mu$s, indicating that the samples in the second ensemble effectively model independent realizations.

\clearpage
\newpage
\section{Supplementary movie}
\subsection{Diffusion flame: Original data and reconstructed data after encoding}
The field variables in the raw data and those reconstructed by the autoencoder (Fig. 4f) are compared.
\subsection{Three-dimensional rendering of the model combustor}
CH$_4$ (blue) and the kernel (orange) are rendered for a representative combustor simulation with successful ignition.
\subsection{Cross-section of the representative simulation samples of the model combustor}
The temperature, CH$_4$ mole fraction, and the pressure on the cross-plane are shown for four representative samples in the first ensemble. The laser amplitudes of the samples are (from the top) 2.7 mJ, 11.1 mJ, 15.7 mJ, and 37.3 mJ.

\bibliographystyle{unsrt}